\documentclass[fleqn,usenatbib]{mnras}

\usepackage{newtxtext,newtxmath}

\usepackage[T1]{fontenc}
\usepackage{ae,aecompl}

\usepackage{graphicx}	
\usepackage{amsmath}	
\usepackage{amssymb}	
\usepackage{lineno}

\title[XCS: A new search for the 3.5 keV feature]{The \textit{XMM} Cluster Survey: new evidence for the 3.5 keV feature in clusters is inconsistent with a dark matter origin}

\author[S. Bhargava et al.]{S. Bhargava$^{1,
2}$\thanks{s.bhargava@sussex.ac.uk},
P. A. Giles$^{1}$\thanks{p.a.giles@sussex.ac.uk},
A. K. Romer$^{1}$,
T. Jeltema$^{2}$,
J. Mayers$^{1}$,
A. Bermeo$^{1}$,
\newauthor
M. Hilton$^{3,4}$,
R. Wilkinson$^{1}$,
C. Vergara$^{1}$,
C. A. Collins$^{5}$,
M. Manolopoulou$^{6}$,
\newauthor
P. J. Rooney$^{1}$,
S. Rosborough$^{1,7}$,
K. Sabirli$^{1}$,
J. P. Stott$^{8}$,
E. Swann$^{1,9}$,
P. T. P. Viana$^{10, 11}$
\\
$^{1}$Department of Physics and Astronomy, University of Sussex, Falmer, Brighton BN1 9QH, UK\\
$^{2}$Santa Cruz Institute for Particle Physics, University of California, Santa Cruz, 1156 High St, Santa Cruz, CA 95064, USA\\
$^{3}$Astrophysics Research Centre, University of KwaZulu-Natal, Westville Campus, Durban 4041, SA\\
$^{4}$School of Mathematics, Statistics, and Computer Science, University of KwaZulu-Natal, Westville Campus, Durban 4041, SA\\
$^{5}$Astrophysics Research Institute, Liverpool John Moores University, Liverpool Science Park, 146 Brownlow Hill, Liverpool L3 5RF, UK\\
$^{6}$Institute for Astronomy, University of Edinburgh, Royal Observatory, Blackford Hill, Edinburgh EH9 3HJ, UK\\
$^{7}$Rochester Institute of Technology, 1 Lomb Memorial Dr, Rochester, NY 14623, United States\\
$^{8}$Department of Physics, Lancaster University, Lancaster LA1 4YB, UK\\
$^{9}$Institute of Cosmology and Gravitation, Dennis Sciama Building, Burnaby Road, Portsmouth PO1 3FX, UK\\
$^{10}$Instituto de Astrof\'isica e Ci\^{e}ncias do Espa\c co, Universidade do Porto, CAUP, Rua das Estrelas, 4150-762 Porto, Portugal\\
$^{11}$Departamento de F\'isica e Astronomia, Faculdade de Ci\^{e}ncias, Universidade do Porto, Rua do Campo Alegre, 687, 4169-007 Porto, Portugal \\
}

\date{Accepted 2020 June 18. Received 2020 June 17; in original form 2020 May 05
}

\pubyear{2020}

\begin{document}
\label{firstpage}
\pagerange{\pageref{firstpage}--\pageref{lastpage}}
\maketitle

\begin{abstract}
There have been several reports of a detection of an unexplained excess of X-ray emission at $\simeq$ 3.5~keV in astrophysical systems. One interpretation of this excess is the decay of sterile neutrino dark matter. The most influential study to date analysed 73 clusters observed by the \textit{XMM-Newton} satellite. We explore evidence for a $\simeq$ 3.5~keV excess in the \textit{XMM}-PN  spectra of 117 redMaPPer galaxy clusters ($0.1 < z < 0.6$). In our analysis of individual spectra, we identify three systems with an excess of flux at $\simeq$ 3.5~keV. In one case (XCS J0003.3+0204) this excess may result from a discrete emission line. None of these systems are the most dark matter dominated in our sample. We group the remaining 114 clusters into four temperature ($T_{\rm X}$) bins to search for an increase in $\simeq$ 3.5~keV flux excess with $T_{\rm X}$ - a reliable tracer of halo mass. However, we do not find evidence of a significant excess in flux at $\simeq$ 3.5~keV in any $T_{\rm X}$ bins. To maximise sensitivity to a potentially weak dark matter decay feature at $\simeq$ 3.5~keV, we jointly fit 114 clusters. Again, no significant excess is found at $\simeq$ 3.5~keV. We estimate the upper limit of an undetected emission line at $\simeq$ 3.5~keV to be $2.41 \times 10^{-6}$ photons cm$^{-2}$ s$^{-1}$, corresponding to a mixing angle of $\sin^2(2\theta)=4.4 \times 10^{-11}$, lower than previous estimates from cluster studies. We conclude that a flux excess at $\simeq$ 3.5~keV is not a ubiquitous feature in clusters and therefore unlikely to originate from sterile neutrino dark matter decay.


\end{abstract}

\begin{keywords}
X-rays: galaxies: clusters; galaxies: clusters: intracluster medium; line: identification; dark matter
\end{keywords}

\newcounter{RMDone}
\setcounter{RMDone}{0}
\def\RM{redMaPPer }



\section{Introduction}

Galaxy clusters, the largest gravitationally collapsed objects in the universe, are vast assemblages of dark matter and hot gas, making them invaluable probes of both cosmology and astrophysics. The dark matter content is estimated to be roughly 85 percent of the total cluster's mass, encompassing each member galaxy's dark matter halo in addition to a larger cluster halo. Hot, energetic gas is found between the cluster member galaxies, forming the intracluster medium (ICM). The ICM is a plasma of predominantly ionized hydrogen and helium, which emits X-ray radiation via the thermal bremsstrahlung process. The gas is additionally enriched with heavier ions which can be detected via their emission lines in the cluster's X-ray spectrum, e.g. O, Ne, Mg, Si, S, Ar, Ca, Fe, and Ni \citep[see][for a review]{Bohringer2010}. The spectral properties of clusters have been studied for several decades, so it came as a surprise when a previously unknown feature at $\simeq 3.5$ keV was reported from an analysis of stacked and individual \textit{XMM-Newton} spectra of 73 galaxy clusters \citep[][hereafter B14, $0.01 < z < 0.35$]{Bulbul2014}.

One interpretation of the feature found by B14 (and in subsequent analyses, see below) is the decay of dark matter in the form of a resonantly produced `sterile' neutrino with mass $m_s\simeq7.1$ keV \citep[][]{Abazajian2001}. Such a particle would have an associated decay mode which results in the two-body state of an active neutrino and a photon with an energy $E=m_s/2$ \citep{PalWolfenstein1982}. Based on the Tremaine-Gunn bound \citep{TremaineGunn}, sterile neutrino dark matter is expected to be in the keV mass range, with $m_s \geq 400$ eV and a lifetime of $\tau_s \geq 10^{24}$ sec \citep{Boyarsky2009}. While the sterile neutrino hypothesis has attracted a lot of attention in the literature, alternative explanations for the $\simeq 3.5$ keV excess have been proposed. These include an interaction of axion-like particles (ALP) with photons \citep[e.g.][]{Berg2017}; elemental plasma transitions whose precise energy are not resolvable by current X-ray telescopes \citep{JeltemaProfumo2015}; and charge exchange processes due to sulphur ions \citep[e.g.][]{Gu2015ChargeExchange,Shah2016ChargeExchange}.

In \cite{Bulbul2016}, the B14 team followed up their original study with an analysis of stacked \textit{Suzaku} spectra of 47 clusters ($0.01 < z < 0.45$). From the \textit{Suzaku} data, they reported another detection of a $\simeq 3.5$ keV feature, albeit at lower significance. There have also been reports of a $\simeq 3.5$ keV excess in the spectra of individual galaxy clusters. The B14 study included an analysis of the Perseus cluster ($z \simeq 0.02$), in which a feature with an anomalously high flux (compared to stacked clusters) was detected. They suggested that contamination from the Ar XVII dielectronic recombination line at 3.62 keV was likely responsible for this anomalously high value. Another study by \citet{Urban2015} using \textit{Suzaku} observations similarly found evidence of a $\simeq 3.5$ keV excess in both the core and the outskirts of the Perseus cluster. However, \citet{Urban2015} concluded that the flux ratio between the core and the outskirts was incompatible with a dark matter interpretation. Moreover, recent high resolution observations of the Perseus cluster by the \textit{Hitomi} satellite failed to find any evidence for a discrete emission line at $\simeq 3.5$ keV \citep[e.g.][]{HitomiPerseus2017,Tamura2019}. The \cite{Urban2015} study also included individual analyses of three other nearby X-ray bright clusters (Coma, Virgo and Ophiuchus) but found no evidence of an excess in their respective spectra. Two of these clusters - Coma and Ophiuchus - were also stacked in B14 along with Centaurus, yielding no evidence of a line. 

In addition to the cluster studies outlined above, there have been several searches for a flux excess at $\simeq 3.5$ keV in the X-ray spectra of other types of astrophysical systems. For example, \citet{Boyarsky2014} reported a detection of a feature consistent with a $\simeq 3.5$ keV emission line in the spectrum of the Andromeda galaxy (M31), although a subsequent analysis found its observed spectrum to be consistent with no excess at $\simeq 3.5$ keV \citep{JeltemaProfumo2015}. \citet{BoyarskyGC} reported a detection of a feature consistent with a $\simeq 3.5$ keV emission line in observations of the Galactic Centre (GC). However, \citet{JeltemaProfumo2015} had previously interpreted the signal in this region of the GC spectrum as the result of plasma emission lines. Furthermore, an analysis of \textit{Chandra} observations of the GC reported no detection of a $\simeq 3.5$ keV feature \citep{Riemer_Sorensen_2016}. Other searches for a $\simeq 3.5$ keV feature have analysed the spectra of: the Galactic bulge \citep{Hofmann2019}; individual galaxies \citep{StackedGalaxiesAnderson,DracoJeltemaandProfumo,Ruchayskiy2016}; galaxy stacks \citep{StackedDSphsMalyshev}; and X-ray blank sky observations of the Milky Way \citep{DessertBlankSky2018}. 

In this work, we revisit the seminal work of B14 by searching for a $\simeq 3.5$ keV flux excess in \textit{XMM-Newton} cluster spectra. Our cluster sample is larger than its precursor, 117 clusters compared to 73 studied in B14, allowing us to examine the detectability of a potential dark matter decay line at $\simeq 3.5$ keV as a function of X-ray temperature ($T_{\textrm{X}}$), and hence, dark matter halo mass. If a $\simeq 3.5$ keV line is detected, and its flux increases with $T_{\textrm{X}}$, then that would lend weight to a dark matter interpretation. However, if the flux weakens with $T_{\textrm{X}}$, then an astrophysical origin would be more likely, since prominent emission lines in the $3-4$ keV region, e.g. K XVII, Ar XVII, K XIX, weaken with plasma temperature (see Figure 4 and Figure 8 in B14 and \citet{Urban2015} respectively). 

In Section \ref{sec:sample}, we describe the sample selection. In Section \ref{sec:methodology}, we describe the method used to test for the presence of a $\simeq 3.5$ keV flux excess. In Section \ref{sec:results}, we present our results. Validation checks and implications of our results are detailed in Section \ref{sec:discussion}. We state our conclusions in Section \ref{sec:conclusions}. Throughout the paper, the parameters $R_{500}$ and $M_{500}$ are calculated with respect to the critical density ($\rho_c$) at the measured cluster redshift. We assume a flat $\Lambda$CDM cosmology with $H_0 = 70$ km s$^{-1}$ Mpc$^{-1}$, $\Omega_{M} = 0.3$ and $\Omega_{\Lambda} = 0.7$. Unless otherwise stated, we use the 68$\%$ $(1\sigma)$ confidence level for all quoted errors in this analysis. 

\section{The cluster sample}
\label{sec:sample}

For this study, we use a subset of clusters drawn from a new sample of 482 clusters presented in a companion paper (Giles et al. (in prep), G20 hereafter). The G20 sample was developed by crossmatching the redMaPPer (hereafter RM) SDSS DR8 cluster catalogue \citep[SDSSRM,][]{RM} with the public \textit{XMM-Newton} data archive, where the X-ray data were processed as part of the \textit{XMM} Cluster Survey \citep[XCS,][]{Romer2001}. 

\begin{figure} 
    \begin{centering}
    \includegraphics[width=0.45\textwidth]{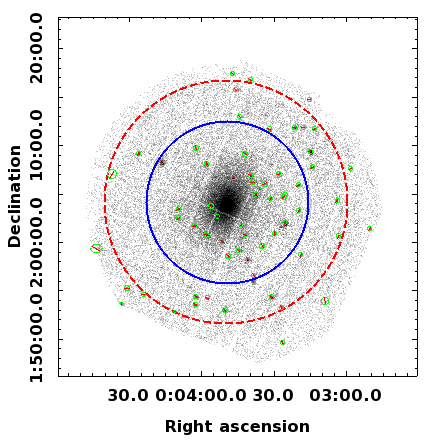}
    \caption{\textit{XMM-Newton} image of XCS J0003.3+0204 in the 0.5--2.0~keV band. The source region is defined by the blue circle. The red dashed-circle defines the background region. Point sources are circled in green and excluded from the spectrum. The cluster image is a composite of PN and MOS observations (ObsID: 0201900101) with an associated redshift from the SDSSRM catalogue of $z_{\rm phot}^{\rm RM}=0.11$.}
    \label{fig:RXCimage}
    \end{centering}
\end{figure}

Of the 482 clusters in the G20 sample, 346 have reliable X-ray temperature measurements (i.e. $\Delta T_{\rm X}/T_{\rm X}<0.25$) in the the redshift range $0.1 < z_{\rm phot}^{\rm RM} < 0.6$. The temperatures in G20 are measured from all available \textit{XMM} data for a given cluster, i.e. from the three cameras on board \textit{XMM} -- PN, MOS1 and MOS2 -- and all available observations if the cluster has been exposed multiple times. The cluster spectra are extracted in the $0.3-7.9$ keV energy band using circular source apertures with a radius of $R_{500}$, and annular background regions spanning $1.05R_{500}$ to $1.5R_{500}$ (see Fig.~\ref{fig:RXCimage} for an example). The $R_{500}$ values are estimated following an iterative method using the $R_{500}-T_{\rm X}$ scaling relation from \cite{Arnaud2005}. During the spectral fitting, three of the five parameters are frozen: the redshift at the value given in the SDSSRM catalogue (a photometric estimate with $|\Delta z|/(1+z) \simeq 0.006$ where $\Delta z = z_{\textrm{phot}}^{\rm RM} - z_{\textrm{spec}}$), the metal abundance at $Z_{\odot}=0.3$ \citep[a value typical for X-ray clusters, see][]{2012ARA&A..50..353K}, and hydrogen column density, $n_{\textrm{H}}$, at the value obtained from the HI4PI survey \citep{HI4PI2016}. The remaining two parameters, $T_{\rm X}$ and normalisation, are left free. More detail about the crossmatching process between SDSSRM and XCS, X-ray spectral analysis, and quality control methods can be found in G20.

\begin{figure}
    \includegraphics[width=0.5\textwidth]{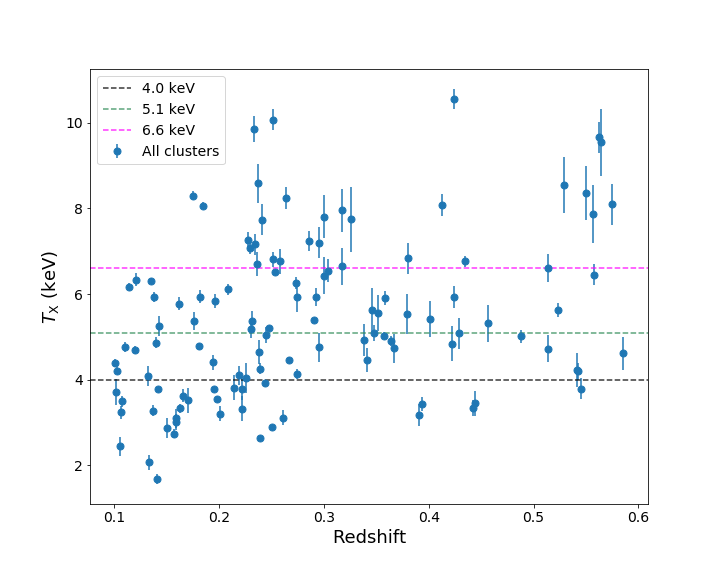}
    \caption[]{The X-ray temperature and redshift distributions of the 117 galaxy clusters used in this study. The dashed lines indicate the boundaries used to define each of the four temperature bins (see Table~\ref{tab:props}).}
    \label{fig:rm-redshift-distribution}
\end{figure}

For the purposes of the current study, we require high fidelity X-ray spectra. Therefore, rather than using all 346 clusters, we apply additional quality controls, detailed as follows. First, we re-derive $T_{\rm X}$ and $\Delta T_{\rm X}$ values using only data from the PN camera and, if multiple observations are available, only from the longest cleaned exposure of a given cluster. We compute the cumulative cleaned exposure time from each PN observation in our sample, obtaining a total of 2.7~Ms of good exposure\footnote{compared to 2.0~Ms of PN data in B14.}. We also calculate associated $0.3-7.9$ keV PN-only signal-to-noise ratios. After applying an upper limit of $\Delta T_{\rm X}/T_{\rm X}|^{\rm PN}=0.1$, and a lower limit signal-to-noise ratio of SNR$=25$, 117 clusters remain. Relevant properties of the 117 clusters are presented in Table~\ref{tab:data}, and Figure~\ref{fig:rm-redshift-distribution} shows the sample distribution of X-ray temperature and RM determined redshifts. Only 13 of these clusters are found to be in common with the B14 analysis (these are indicated in Appendix~\ref{tab:data}).

\begin{table*}
    \begin{centering}\scriptsize
\caption[]{\small{Properties of the cluster sample according to binned X-ray temperature.}\label{tab:props}}
\vspace{-2mm}
\begin{tabular}{cccccccc}
\hline\hline
Bin number & ${T_{\rm X}}$ bin & No. of clusters & $T_{\textrm{X}}$ average & $M_\text{DM}^\text{proj}$ average & Fitted abundance & \multicolumn{2}{c}{SNR average} \\
\cline{7-8}
 & (keV) & & (keV) & (10$^{14}$ $M_\odot$) & $Z_\odot$ & 0.3 -- 7.9~keV & 3.0 -- 4.0~keV \\
\hline
1 & $\leq 4$ & 30 & 3.24 & 1.88 & 0.24 & 89.6 & 14.7 \\       
2 & $4 - 5.1$ & 29 & 4.60 & 3.26 & 0.34 & 118.9 & 22.8 \\        
3 & $5.1 - 6.6$ & 28 & 5.82 & 4.92 & 0.20 & 179.0 & 37.1 \\       
4 & $\geq 6.6$ & 30 & 7.89 & 8.07 & 0.29 & 163.8 & 36.5 \\
\hline
\end{tabular}
\begin{flushleft}
    \small{{\bf Notes}. Column (1): Bin number; Column (2): temperature range of the bin; Column (3): number of clusters in bin; Column (4): average temperature of bin; Column (5) average projected dark matter mass ($M_\text{DM}^\text{proj}$); Column (6): fitted abundance (in units of solar metallicity); Column (7): average SNR in the 0.4 -- 7.9~keV band; Column (8): average SNR in the 3.0 -- 4.0~keV band.}
\end{flushleft}
    \end{centering}
\end{table*}

\section{Methodology}
\label{sec:methodology}

\subsection{Blueshifting to the rest frame}
\label{subsec:blueshifting}

Before carrying out spectral fits (described in Sect.~\ref{subsec:jointfitting}), the spectra are blueshifted (i.e. so that $z_{\rm{effective}}=0$). This is not strictly necessary when examining individual clusters, but is required when performing joint fits. For joint fits, the blueshifting process has the additional advantage of `smearing out' any redshift-independent instrumental artefacts that could be mistaken for astrophysical emission lines.

The format of a source spectrum measured by the detector is a list of photon counts as a function of channel number. The associated cluster response matrix file ({\tt RMF}) and ancillary response file ({\tt ARF}) contain the energy ranges corresponding to the source spectrum channels. Each cluster spectrum is blueshifted by rescaling the upper and lower energy bounds for each photon channel by a factor of $1+z$. This shifts the number of photons associated with each energy according to the observed redshift of the cluster. Because the source and background spectra both rely on the {\tt ARF} and {\tt RMF}, modifications to both spectra are required to ensure consistency. We present a validation check of the blueshifting technique in Section~\ref{subsec:blueshiftingtests}. We note that our approach to blueshifting is the same as that used in B14.

\subsection{Spectral fitting}
\label{subsec:jointfitting}

We have carried out three separate but related tests on the cluster spectra: the first is on the 117 clusters separately (Sect.~\ref{subsec:resultsIndiv}) to determine any outliers with excess flux at $\simeq 3.5$ keV. The second is a joint fit to clusters binned into four different temperature bins (Sect.~\ref{subsec:resultsTbin}, with and without outliers). The third is a joint fit to the whole sample (minus the outliers, see Sect.~\ref{subsec:resultsAll}). Each test is progressively more sensitive to the existence of a dark matter decay spectral feature. The second test also allows us to search for a potential mass dependence of a $\simeq$ 3.5 keV feature, because $T_{\textrm{X}}$ is a robust tracer of the underlying dark matter mass. Hence, evidence of an increase in a $\simeq$ 3.5~keV flux excess with $T_{\textrm{X}}$ would give firm support to the dark matter interpretation (and vice versa). For each test, we carry out a fit to a fiducial model (`model A': {\tt tbabs $\times$ apec}) and then compare the goodness of fit to a model that includes an addition emission line component (`model B': {\tt tbabs $\times$ (apec + weight $\times$ Gaussian)}) to mimic a dark matter decay feature. The fitting is performed using {\sc xspec} version {\sc 12.10.1f} \citep{XSPEC}, {\sc apec} version 3.0.9, and solar abundances based on \cite{AndersandGrevesse}, using the {\sc xspec} {\tt cstat} statistic.

There are five parameters in model A. Three are frozen during the fit: the $n_{\rm H}$ value, the X-ray temperature (at the $T_{\rm X}^{\rm PN}$ value, see Sect.~\ref{sec:sample}), and the redshift (at $z_{\rm effective}=0$). Two are left free: the {\tt apec} normalisation, and the metal abundance. During joint fits, the abundance is `tied' across all the spectra being examined. This results in an average abundance per fit (see column 5 in Table \ref{tab:props}). For both individual and joint fits, the normalisation of the electron plasma density is fitted separately to each cluster.

There are nine parameters in model B. Five of these are shared with model A and treated in the same way during the fit. The remaining four parameters are associated with the Gaussian component: the central energy, line width, normalisation, and a constant weighting factor ($0<{\tt weight}<1$).  The central energy is frozen at a value iterated between $3 - 5$ keV in intervals of 25 eV, i.e. 80 separate fits to model B are run for a given analysis. The line width is fixed at zero to mimic the narrowest possible line emission allowed by the energy resolution of the detector, which is in turn defined by the {\tt ARF} matrix associated with the respective cluster spectrum. The normalisation is a free parameter but, like the metal abundance, is fitted jointly or `tied,' generating an average fitted value per bin. The weighting factor is an input to the model, and frozen during the fit. Each cluster has a different assigned weight (see Sect.~\ref{sec:weighting}).


We define the parameter $\Delta C$ to quantify the change in the goodness of fit between the two models at a given energy $E$, where $3<E<5$ keV (see above). $\Delta C$ is the difference between the value of the Cash statistic \citep{Cash1979} after fitting for model A and the value after fitting for model B. A positive value of $\Delta C$ indicates that the fit is better for model B. The estimate for the $3\sigma$ threshold (i.e. the value of $\Delta C$ above which is considered a significantly better fit), is calculated based on the probability of exceeding 99$\%$ significance for model B compared to model A, taking into account the fact that model B has one additional degree of freedom.

\subsubsection{Differences to the B14 method}

Our analysis differs from B14 in several ways. Firstly, we implement the {\tt apec} plasma model using the standard approach, i.e. relying on predefined emissivities taken from {\sc Atomdb} \citep{Foster2012} to account for emission lines. B14 alternatively define a line-free {\tt apec} plasma model with 28 Gaussian models added to account for emission lines (though some are later removed to improve convergence of their fits). Next, with respect to photoelectric absorption, we use the {\tt tbabs} cross-sections, whereas B14 adopt the {\tt wabs} values (see Sect.~\ref{subsec:nhvalue}). Our methods also differ in the approach to background subtraction. We use an infield background subtraction method (see Sect.~\ref{sec:sample}). B14 use a composite background model that accounts for contributions from the quiescent particle background, the cosmic X-ray background, solar wind charge exchange, as well as residual contamination from soft protons. Furthermore, we use the $\Delta C$ parameter to assess the change in the goodness of fit between model A and B \citep[following a similar analysis undertaken by][]{Urban2015}, whereas B14 uses a $\chi^2$ approach.

Whilst we fit each spectrum in parallel when performing joint fits, B14 stack their data into a composite spectrum first. The advantage of our method is that it allows us to explore the influence of individual spectra on the joint fit (see Sect. \ref{subsec:jointfitting}). Moreover, in our study, we have focused on {\em XMM}-PN data, whereas B14 also fitted to {\em XMM}-MOS, as well as analysing the {\em Chandra}-ACIS spectra of Virgo and Perseus. 

Finally, when searching for evidence of a 3.5 keV feature, the energy values in our analysis are frozen in intervals of 25 eV (e.g. 3.5, 3.525, 3.55, 3.575, 3.6 etc.) whereas B14 nominally compute a best fitted value for their energy of an unidentified line in their stacked spectrum. However, we note that out of the 14 fits in their study, best fitted values are only computed for the full {\em XMM} PN and MOS samples. Stacked spectra consisting of fewer clusters subsequently assume a fixed energy at 3.57 keV. Similarly, for the {\em Chandra} ACIS spectra, a best fitted energy is computed for Perseus, and subsequently frozen at 3.56 keV in the {\em Chandra} spectrum of Virgo.

\subsubsection{Dark matter flux and weighting}
\label{sec:weighting}

If a flux excess (over the fiducial model A) originates from dark matter decay, then for a given cluster, we would expect the flux to increase with the projected dark matter mass in the \textit{XMM} FOV, $M_\text{DM}^\text{proj}$, but to decrease with cluster redshift, $z$. To account for this, $M_\text{DM}^\text{proj}$ and $z$ dependent weights are applied during the joint spectral fits. The $M_\text{DM}^\text{proj}$ values are defined within a radius $R_\text{ext} = R_{500}$, i.e. the same extraction aperture as the PN spectrum (stated in Sect. \ref{sec:sample}). The total masses for the clusters are estimated by applying the $M_{500}-T_{\rm X}$ scaling relation described in \cite{Arnaud2005}. These are then corrected for the fact that the dark matter accounts for only $85\%$ of the total mass, and the projected dark matter mass within a $R_{500}$ cylinder is larger than that within a sphere\footnote{This is done following the method in \cite{LokasMamonNFW}, assuming a concentration parameter, $c_{500}=3$, based on the $c_{500}-M_{500}$ scaling relation described in \cite{Vikhlinin2006}}. The projected dark matter mass for each individual cluster is stated in column 4 in Table \ref{tab:data}. The average projected dark matter mass for each temperature bin is stated in column 4 of Table \ref{tab:props}.

For the joint fits, to account for a different dark matter contribution from each cluster, we apply a weighting to each cluster during the fits to model B. We calculate the weighting $w_i$ from each cluster $i$ in a given temperature bin according to
\begin{equation}\label{eq:dmweighting}
    w_\text{i,DM} = \frac{M_\text{i,DM}^{\text{proj}}(<R_{\text{ext}})(1+z_i)}{4\pi d^2_{i,L}},
\end{equation}
where $d_{i,L}$ is the luminosity distance at $z_i$. Before the fitting to model B takes place, the individual cluster weights $w_\text{i,DM}$ are normalised by the largest value in the chosen bin, i.e. one cluster per bin has a $\tt{weight}=1$, while remaining clusters have $0<\tt{weight}<1$. During the fits to individual clusters, the weighting is assigned to unity.

\subsection{Estimation of sterile neutrino mixing angles}

If a measured flux excess is due to dark matter decay, we can estimate a sterile neutrino mixing angle using the Gaussian line normalisation taken from the fit to model B. For this we use the B14 relation between decaying dark matter flux, $F_{\text{DM}}$ and projected dark matter mass,
\begin{equation}\label{eq:mixingangle}
    \begin{centering}
    \begin{split}
    \sin^{2}2\theta = \frac{F_{\text{DM}}}{12.76 \, \text{cm}^{-2}\text{s}^{-1}}
    \left(\frac{10^{14} M\odot}{M^{\text{proj}}_{\text{DM}}}\right) \\
    \left(\frac{d_L}{100 \, \text{Mpc}}\right)^{2}
    \left(\frac{1}{1+z}\right)
    \left(\frac{\text{keV}}{m_s}\right)^4.
    \end{split}
    \end{centering}
\end{equation}

\section{Results}
\label{sec:results}

\subsection{Fits to individual clusters}
\label{subsec:resultsIndiv}

\begin{figure}
    \begin{centering}
    \includegraphics[width=0.45\textwidth]{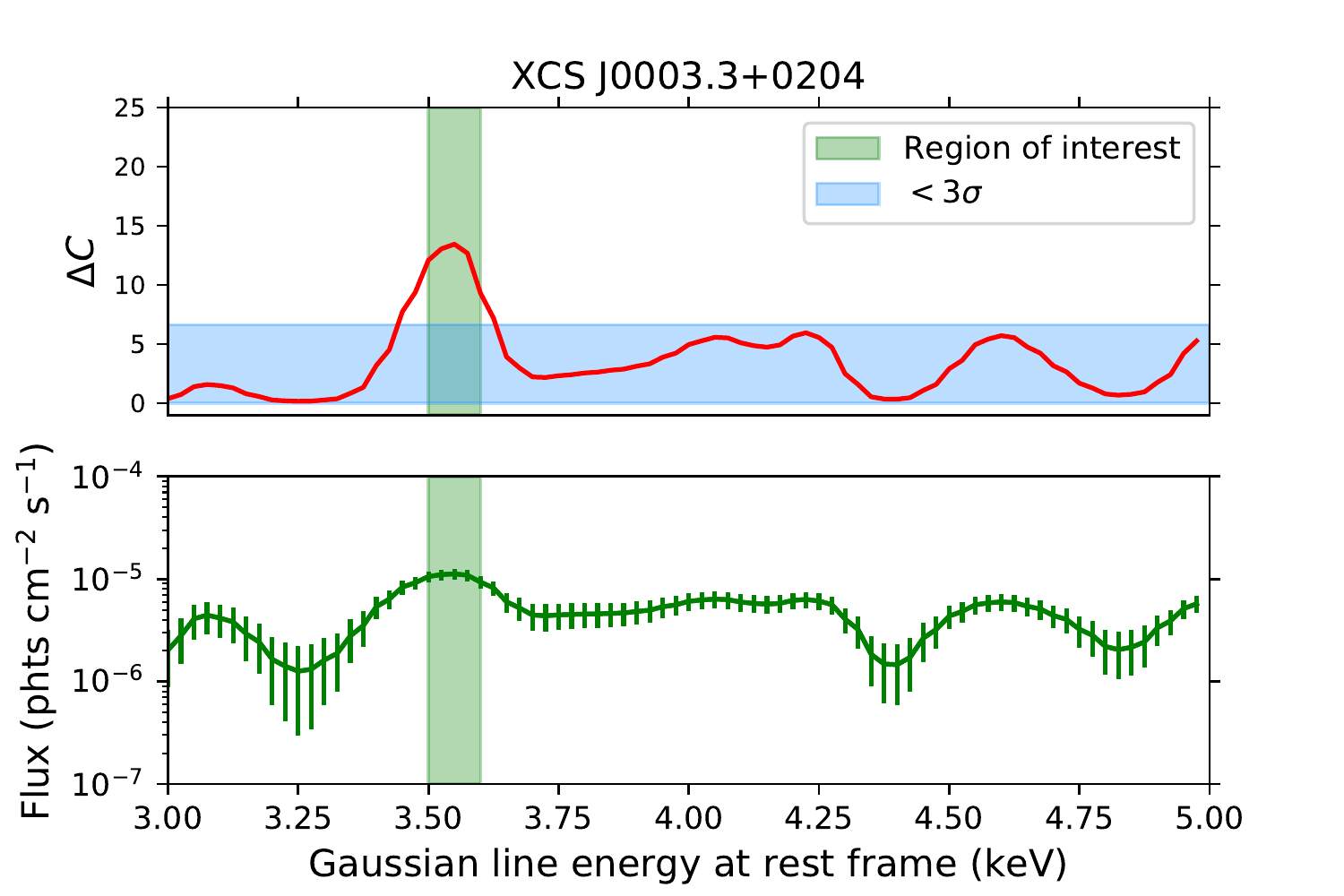}
    \includegraphics[width=0.45\textwidth]{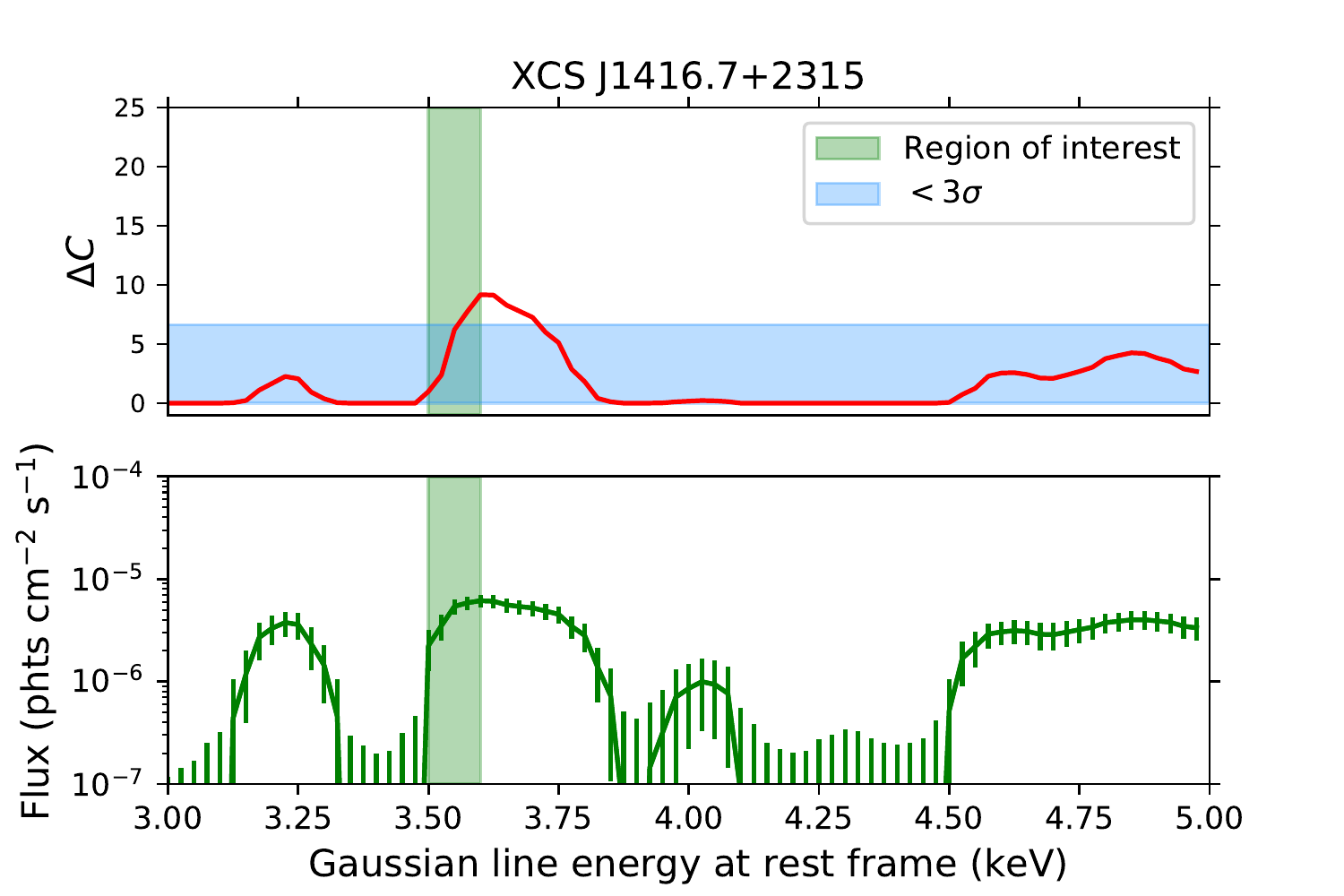}
    \includegraphics[width=0.45\textwidth]{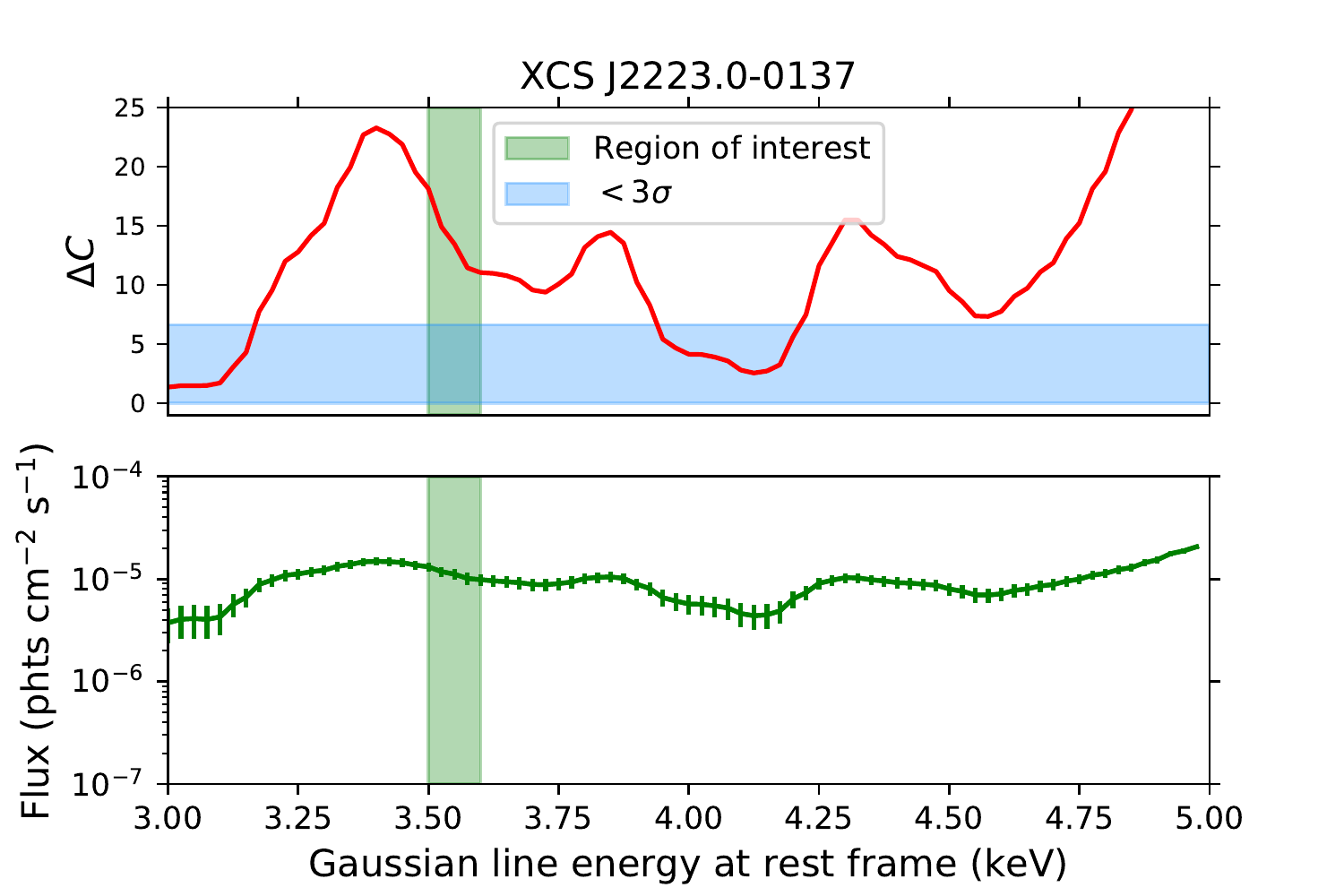}
    \caption{Top panels (red line): The change in fit statistic ($\Delta C$) between model A and model B (see Sect~\ref{subsec:jointfitting}) as a function of energy in the range $3 - 5$ keV. Bottom panels (green line): Fitted normalisation of the Gaussian line and corresponding errors.  The value of $\Delta C$ determines the extent to which model B is a better fit to the data than model A. In each plot, the green shaded region encloses the 3.5 -- 3.6 kev range (where a potential DM signal is expected), defined as the `region of interest.' The light blue shaded region determines a $<$3$\sigma$ detection (see Sect.~\ref{sec:results} for definition). The top, middle and bottom plots refer to the clusters XCS J0003.3+0204, XCS J1416.7+2315, and XCS J2223.0-0137 respectively.}
    \label{fig:individualclusters}
    \end{centering}
\end{figure}

To look for evidence of a $\simeq 3.5$ keV feature in individual cluster observations, as has previously been reported from Perseus \citep[e.g. B14,][]{Urban2015,Franse2016}, we perform a $\Delta C$ analysis on each of the 117 clusters in our sample, using the methodology described in Section \ref{subsec:jointfitting}. We find evidence for a >$3\sigma$ fit improvement at $\simeq 3.5$ keV in three cases: XMMXCS J000349.3+020404.8 (hereafter XCS J0003.3+0204), XMMXCS J141627.7+231523.5 (hereafter XCS J1416.7+2315) and XMMXCS J222353.0-013714.4 (hereafter XCS J2223.0-0137). The results are shown in Figure~\ref{fig:individualclusters}. For each cluster, the top panel shows $\Delta C$ as a function of energy, and the bottom panel shows the corresponding normalisation of the Gaussian line component in units of photons cm$^{-2}$s$^{-1}$. The horizontal blue shaded area in the top panels represents the $< 3\sigma$ region. The vertical green bars in both top and bottom panels span $3.50 - 3.60$ keV, which indicates the expected energy range in which a $\simeq 3.5$ keV line would be detected assuming the appropriate spectral resolution for the instrument. All prior detections of the 3.5 keV feature have quoted a best fit energy firmly within this range, hereafter known as `the region of interest.'

For XCS J0003.3+0204 (XCS J1416.7+2315), the largest fit improvement occurs at 3.55 (3.6) keV, characterised by $\Delta C = 13.4$ (9.17) and a corresponding Gaussian line flux of $1.12_{-0.31}^{+0.31} \times 10^{-5}$ ($6.14_{-1.99}^{+2.02}$ $\times 10^{-6}$) photons cm$^{-2}$ s$^{-1}$. For XCS J2223.0-0137, the maximum value of $\Delta C$ falls below the region of interest, although still exceeds $3\sigma$ therein. Further discussion of XCS J0003.3+0204, XCS J1416.7+2315 and XCS J2223.0-0137 can be found in Sections~\ref{subsec:xcs0003}, \ref{subsec:xcs1416}, and \ref{subsec:xcs2223}, respectively. 

The sterile neutrino mixing angle estimates for XCS J0003.3+0204 and XCS J1416.7+2315 are given in Table~\ref{tab:3.5kevprops}. We do not provide this information for XCS J2223.0-0137 because, from Figure~\ref{fig:individualclusters}, the shape of the flux excess found in this cluster is unlikely to indicate the presence of a discrete emission feature (see Sect. \ref{subsec:xcs2223}). We note that the estimated $\sin^{2}2\theta$ values (of order $\times 10^{-9}$) are significantly larger than those measured by B14, based on the PN-only result for their full sample as well as individual clusters, i.e $4.3^{+1.2}_{-1.0}10^{-11}<\sin^{2}2\theta <1.9\times10^{-10}$ (as quoted in Table 5 of B14). 

\begin{table*}
    \begin{centering}\scriptsize
\caption[]{{\small Measured properties of the 3.5 keV excess, under the interpretation that it is due to a dark matter decay line.}\label{tab:3.5kevprops}}
\vspace{-2mm}
\begin{tabular}{cccccccc}
\hline\hline
Sample & Line energy $(E)$ & $\Delta C$ & Flux & $M_\text{DM}^\text{proj}/d_L^2$ & Mixing angle \\
 & (keV) &  & ($10^{-6}$ photons cm$^{-2}$ s$^{-1}$) & (10$^{10}$ $M_\odot$ Mpc$^{-2}$) & ($\sin^{2}2\theta$) \\
\hline
 XCS J0003.3+0204 & 3.55 & 13.4 & $11.2_{-0.31}^{+0.31}$ & 0.14 & $2.36_{-0.65}^{+0.65} \times 10^{-9}$ \\        
 XCS J1416.7+2315 & 3.6 & 9.17 & $6.14_{-2.02}^{+1.99}$ & 0.05 & $3.78_{-1.23}^{+1.24} \times 10^{-9}$ \\       
 Bin 2 (all 29 clusters) & 3.5 & 11.8 & $4.17_{-1.22}^{+1.22}$ & 0.65 & $1.97_{-0.58}^{+0.58} \times 10^{-10}$ \\   
 114 clusters & - & - & $2.41$ & 1.65 & $4.4 \times 10^{-11}$ \\
\hline
\end{tabular}
    \end{centering}
\end{table*}

\subsection{Joint fits to sub-samples binned by temperature}
\label{subsec:resultsTbin}

\begin{figure}
    \begin{centering}
    \begin{tabular}{c}
       \includegraphics[width=0.4\textwidth]{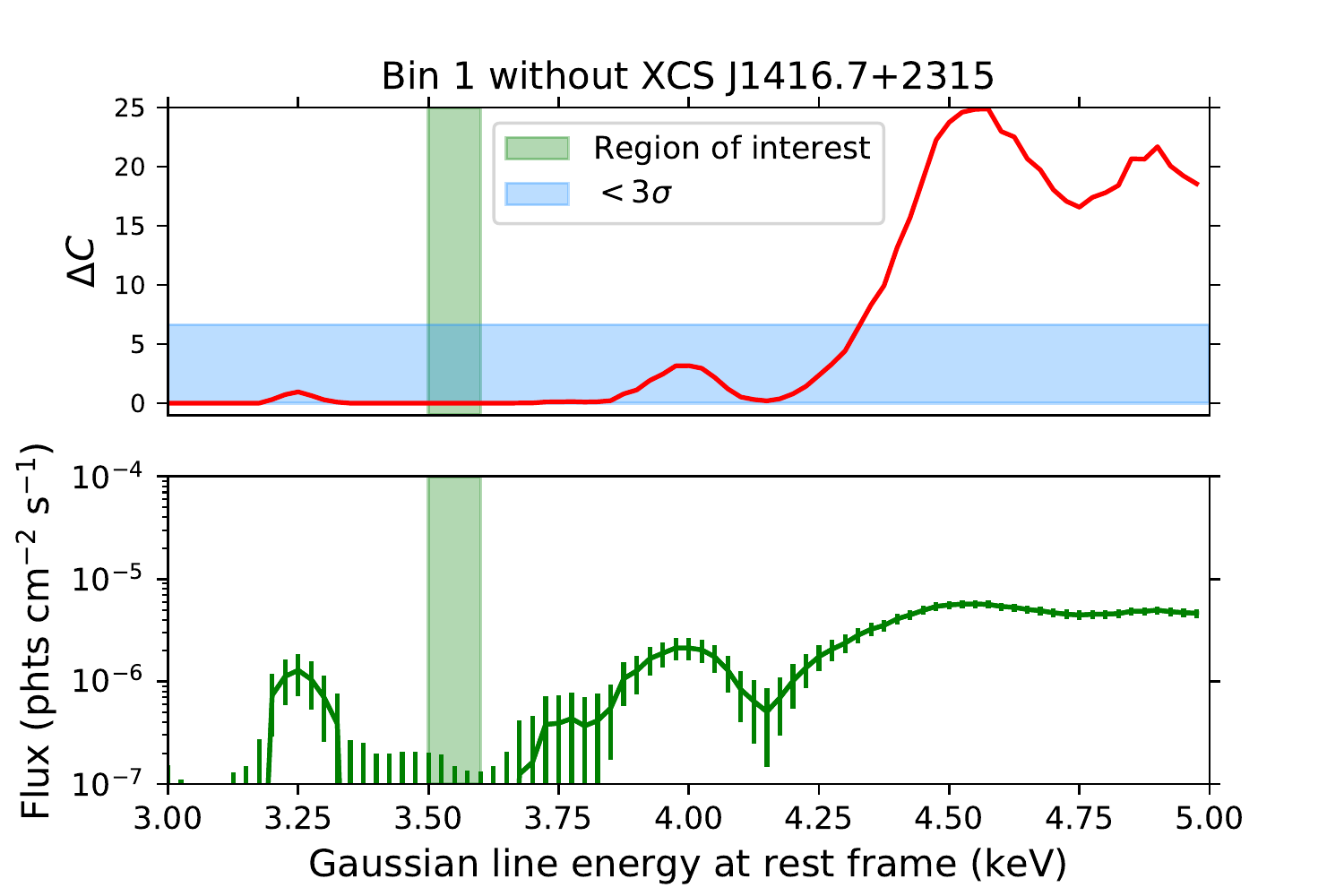}\\
   (a)\\
              \includegraphics[width=0.4\textwidth]{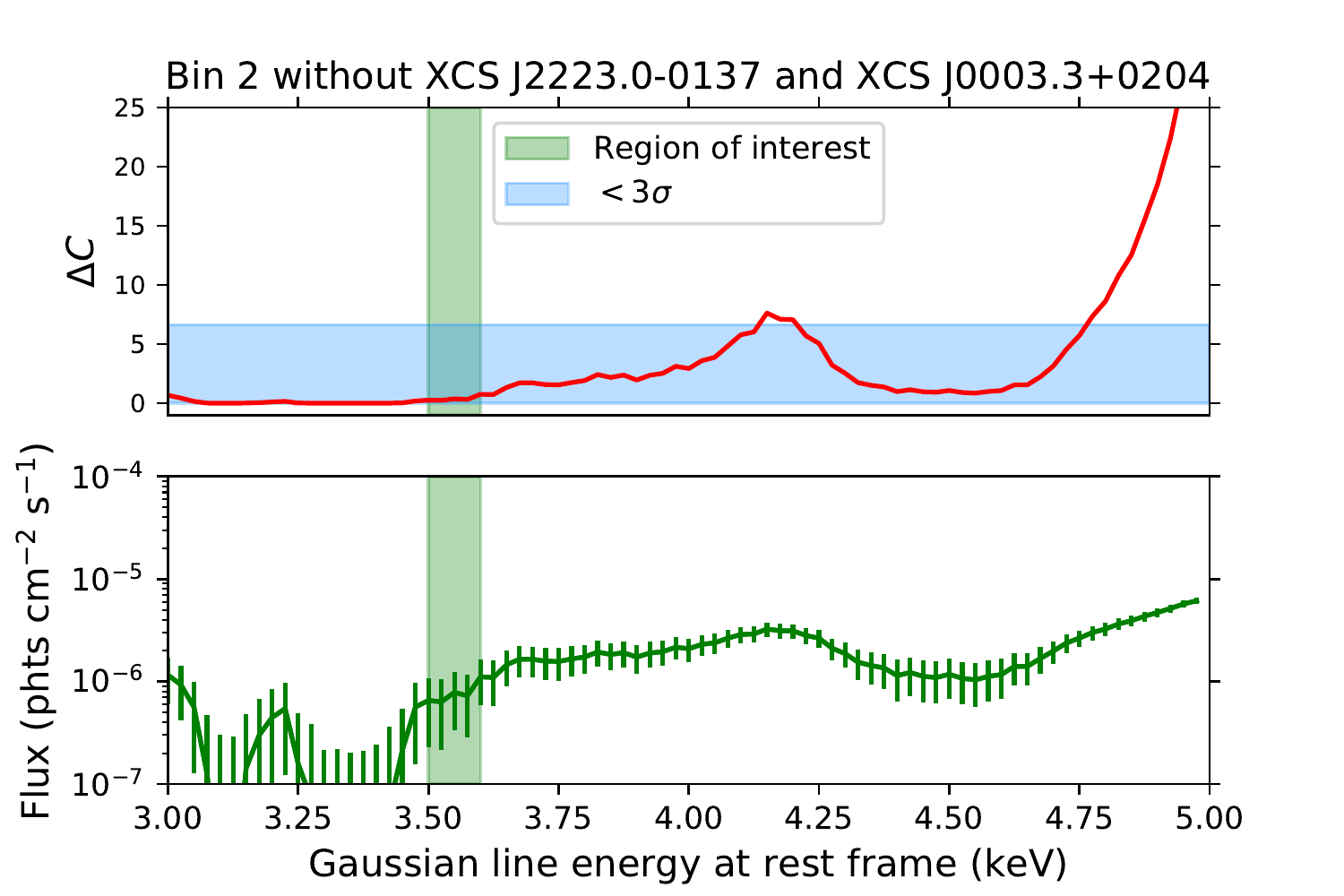} \\
   (b) \\
    \includegraphics[width=0.4\textwidth]{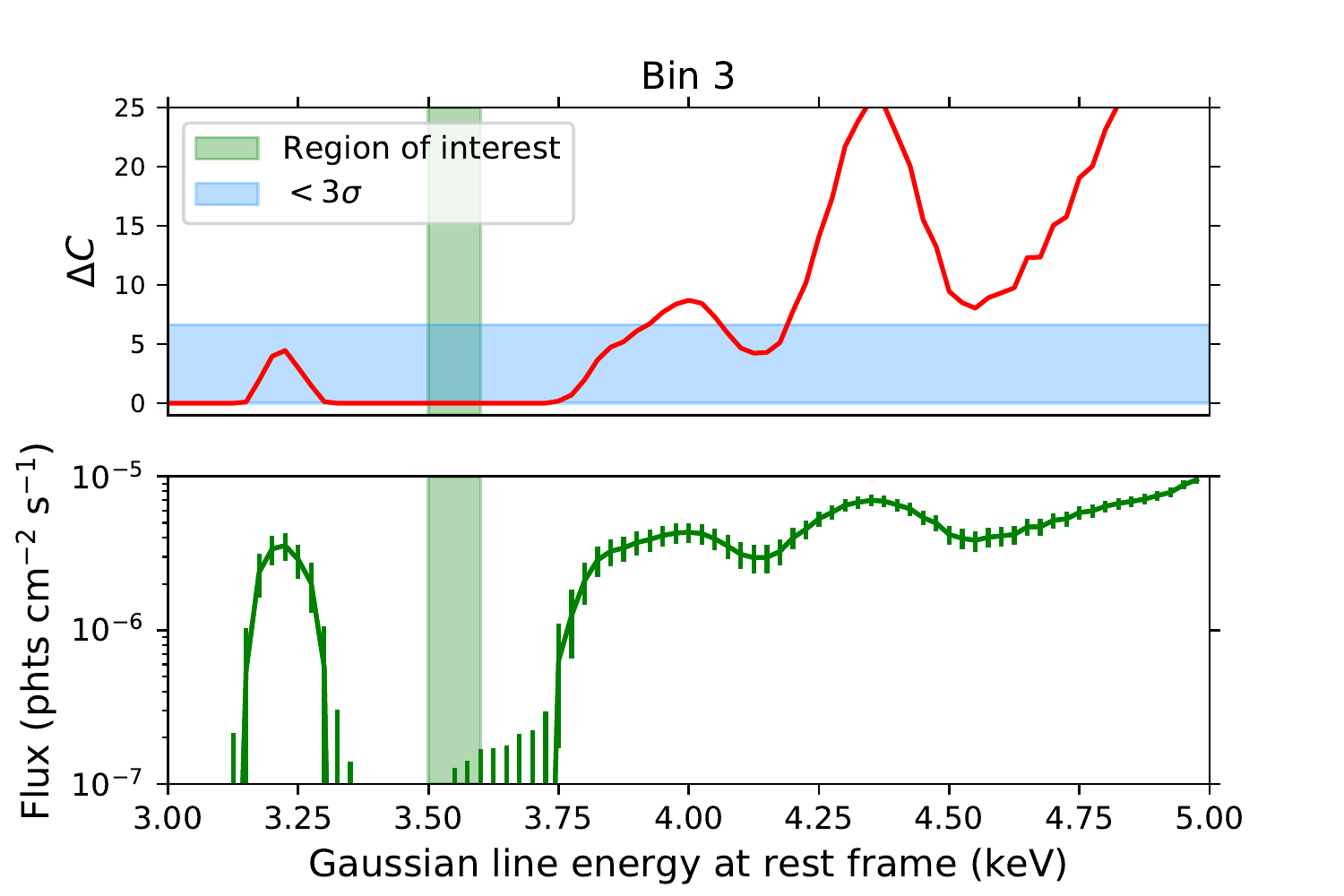}\\
    (c)\\
    \includegraphics[width=0.4\textwidth]{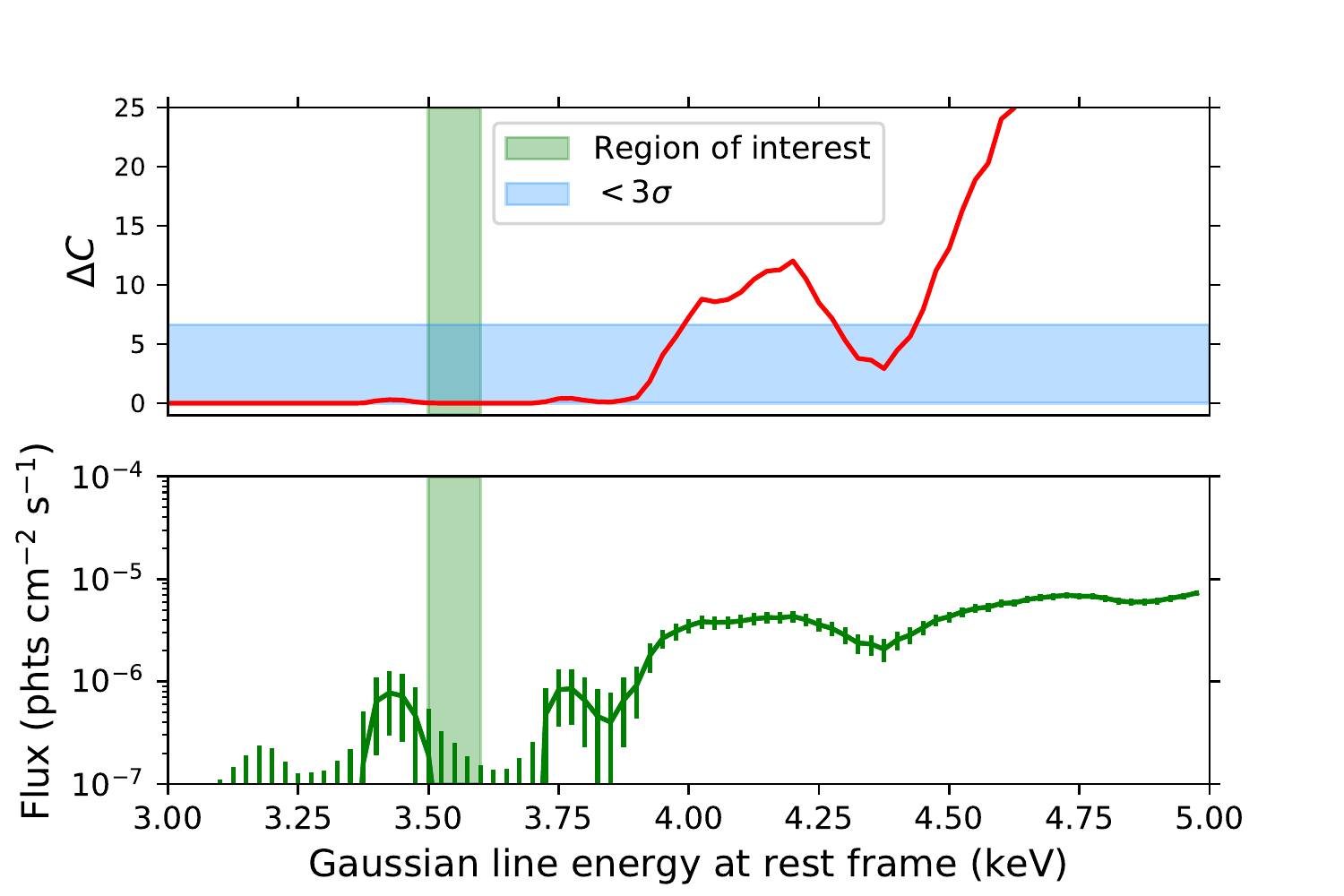} \\
   (d) \\
    \end{tabular}
    \caption{Similar to Figure~\ref{fig:individualclusters}. Results from the binned subsets (see Table~\ref{tab:props}) of clusters excluding those with detected excess at $\simeq$3.5 keV. (a): 29 clusters from bin 1 (i.e. excluding XCS J1416.7+2315). (b): 27 clusters from bin 2 (i.e. excluding XCS J0003.3+0204 and XCS J2223.0-0137). (c) 28 clusters from bin 3. (d) 31 clusters from bin 4.
    }
    \label{fig:tx-all-bin-fits}
    \end{centering}
\end{figure}

To test for a potential temperature dependence of the strength of a $\simeq$ 3.5 keV flux excess, the 117 clusters in the sample are subdivided into four temperature bins: $\leq$4~keV, $4 - 5.1$ keV, $5.1 - 6.6$ keV, $\geq$6.6~keV, containing 30, 29, 28 and 30 clusters respectively. For simplicity, hereafter we refer to these temperature bins as bin 1 ($\leq$4~keV), bin 2 ($4 - 5.1$ keV), bin 3 ($5.1 - 6.6$ keV) and bin 4 ($\geq$6.6~keV). Properties of the bins, averaged according to the number of clusters, can be found in Table \ref{tab:props}. In Figure~\ref{fig:tx-all-bin-fits} we present the results of the $\Delta C$ analysis of each the four temperature bins, after removing the three cases shown in Figure~\ref{fig:individualclusters}.

No significant fit improvement is found in any bin in the region of interest, i.e. the range defined by the vertical green bar.
We note that within the four bins, there are other ranges of $\Delta C$ values that exceed a 3$\sigma$ improvement of model B over model A. These regions correspond to energies where there are known astrophysical lines (e.g. Ar XVII complex with the strongest line at 3.32 keV, Ca XIX at 3.86 keV \& 3.90 keV, Ca XX at 4.1 keV, and Ca XIX at 4.58 keV). Two prominent instrumental lines are also present; the Ti K$\alpha$ at 4.51 keV, and Ti K$\beta$ at 4.93 keV 
\newline
\citep[see][]{DracoJeltemaandProfumo}. Even though the aforementioned plasma lines are included in the latest version of the {\sc apec} model, {\sc apec} does not always correctly predict their relative fluxes as a function of plasma temperature and metal abundance \citep[see e.g.][]{Aharonian2018}, hence fit improvements at the location of known emission lines are not unexpected. Analysis of the Perseus core in \cite{Urban2015} has suggested underestimates of the abundances of elements including Ca XIX and Ti XXII (unresolved lines at 4.97 keV and 4.98 keV), the latter of which is responsible for a high $\Delta C$ value at $\simeq 4.9$ keV in each bin (Fig. \ref{fig:tx-all-bin-fits}).  

\begin{figure}
    \begin{centering}
    \includegraphics[width=0.45\textwidth]{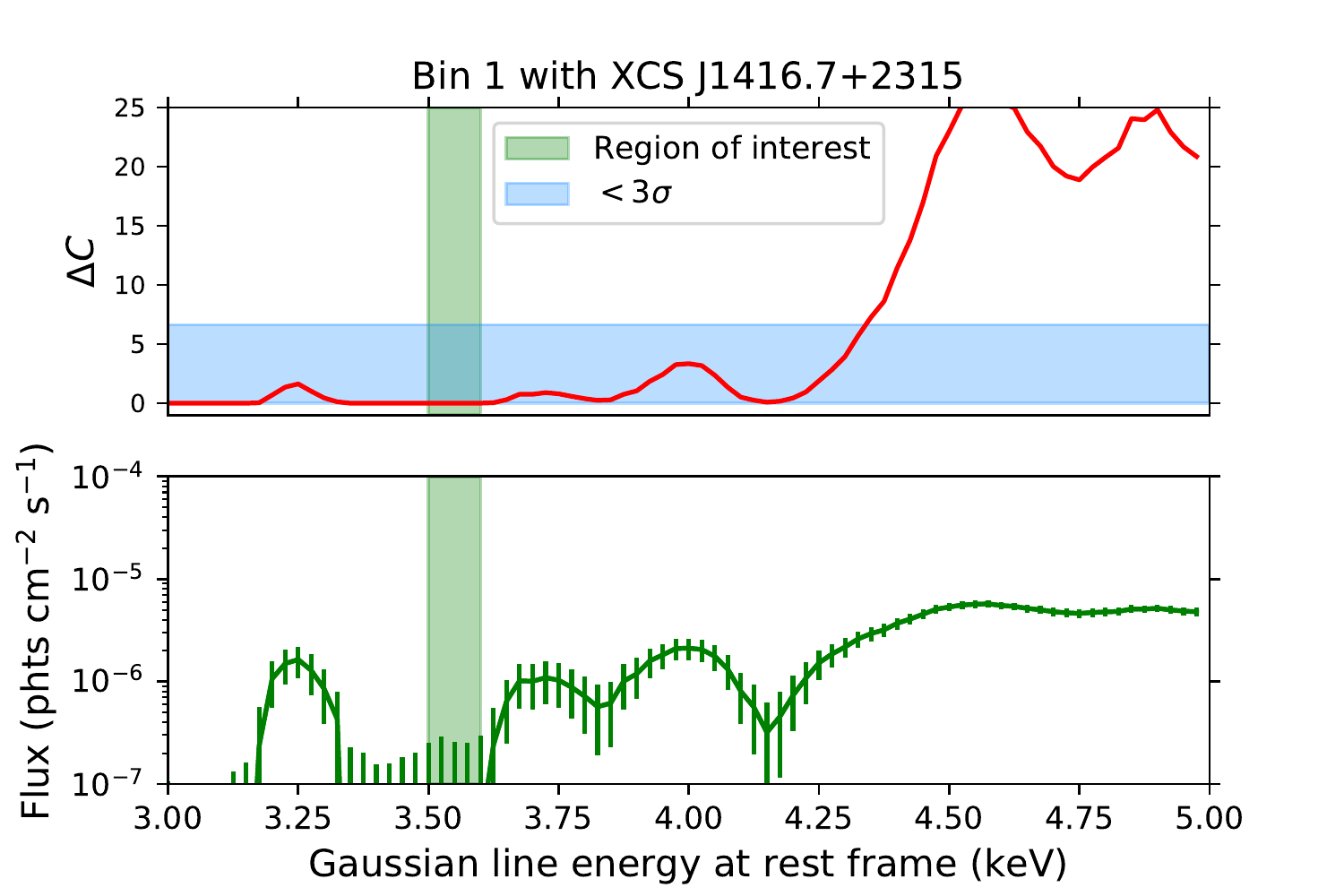} \\
    (a)\\
    \includegraphics[width=0.45\textwidth]{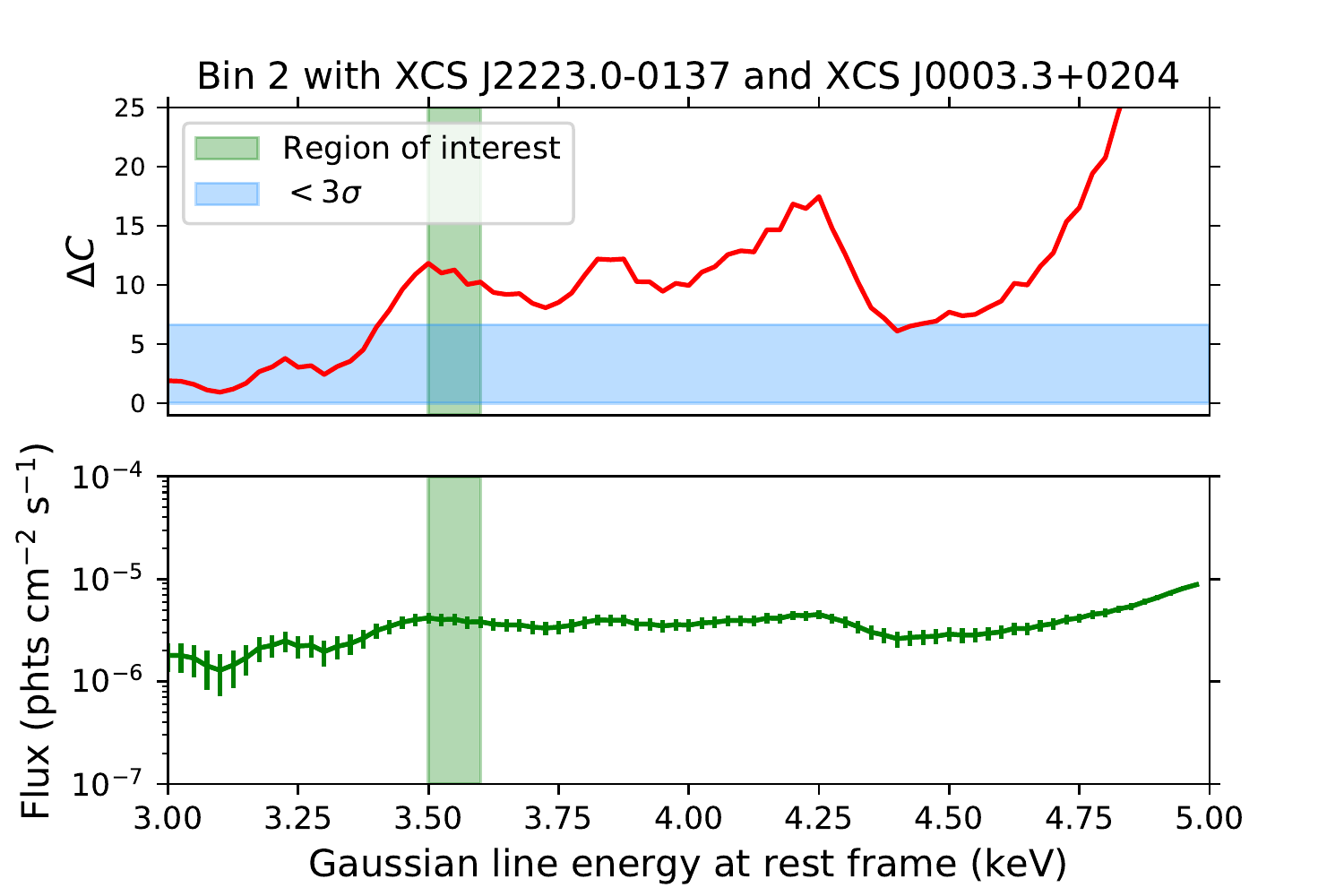} \\
    (b)\\
   \caption{Similar to Figure \ref{fig:tx-all-bin-fits}, but showing the results from the binned subsets of clusters whilst including those with detected excess at $\simeq$3.5 keV (see Table~\ref{tab:props}). (a): All 30 clusters in bin 1 (i.e. including XCS J1416.7+2315). (b): All 29 clusters in bin 2 (i.e. including XCS J0003.3+0204 and XCS J2223.0-0137).}
   \label{fig:tx-all-bin-fits-mc}
    \end{centering}
\end{figure}

For completeness, we repeat the joint analysis of bins 1 and 2 with the clusters featured in Figure~\ref{fig:individualclusters} included (see Fig.~\ref{fig:tx-all-bin-fits-mc}). In Figure \ref{fig:tx-all-bin-fits-mc} (b), when XCS J2223.0-0137 and XCS J0003.3+0204 are included in bin 2, there is now a $>3\sigma$ fit improvement within the region of interest. The maximal fit improvement is found when the central line energy is frozen at $E=3.50$ keV, characterised by a $\Delta C = 11.8$ and corresponding Gaussian line flux of $4.17_{-1.22}^{+1.22} \times 10^{-6}$ photons cm$^{-2}$ s$^{-1}$, corresponding to a mixing angle of $1.97_{-0.58}^{+0.58} \times 10^{-10}$ (see Table \ref{tab:3.5kevprops}).

\begin{figure}
    \includegraphics[width=0.5\textwidth]{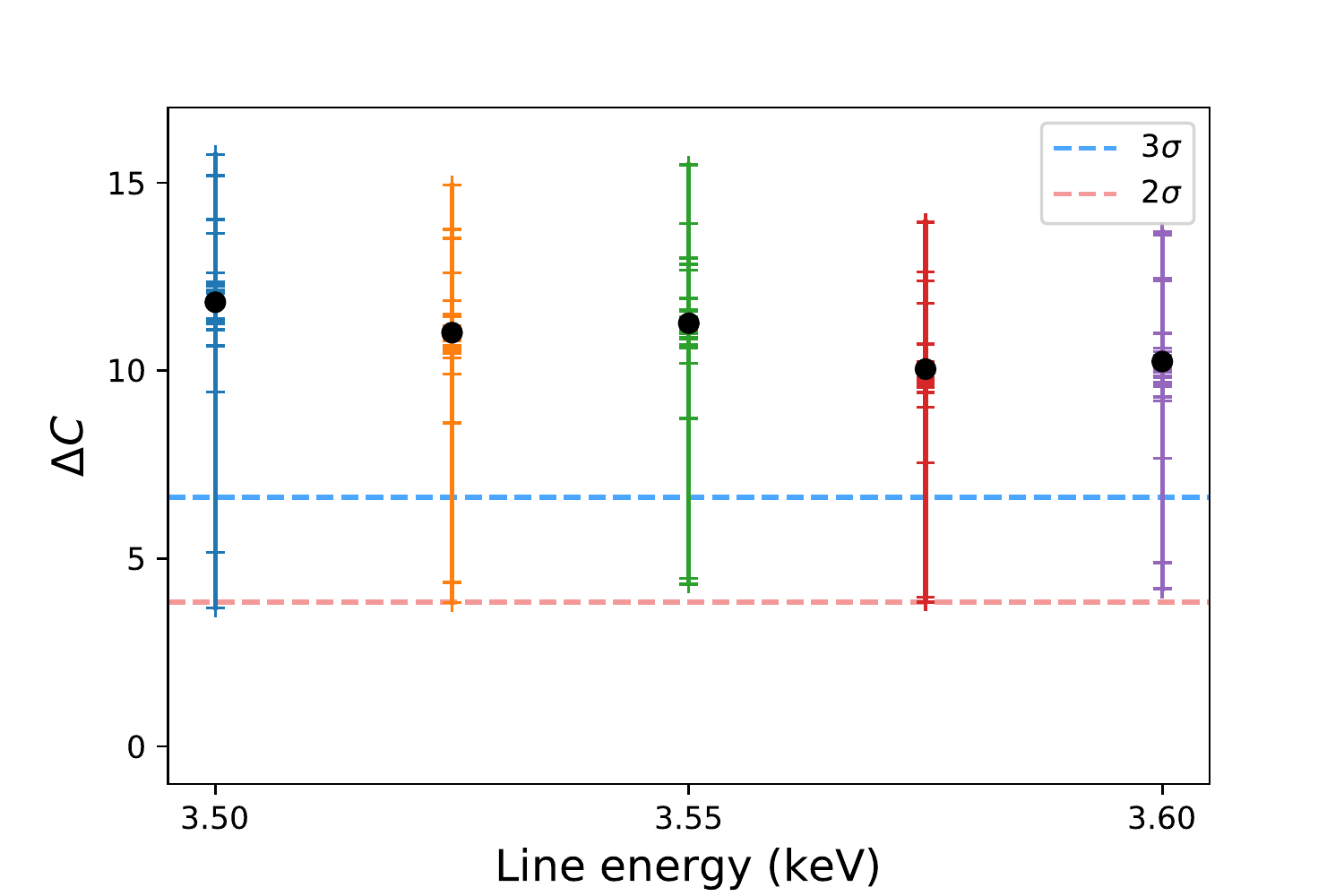}
    \caption{Variation of $\Delta C$ during a jackknife analysis performed at five energy intervals between 3.5 and 3.6 keV in bin 2 (all 29 clusters included). The black data points refer to the value of $\Delta C$ with all clusters included (i.e. the fitted value in Fig. \ref{fig:tx-all-bin-fits-mc}(b)). Each tick-mark refers to the value of $\Delta C$ when a cluster is removed.}
    \label{fig:jackknifedplot}
\end{figure}

To further investigate the influence of individual clusters on the joint fit in each bin, a jackknifing resampling procedure is used: for a temperature bin containing $N$ clusters, we perform $N$ fittings in each bin containing $N-1$ clusters at 5 equally spaced values of the central Gaussian line energy ($3.5<E<3.6$ keV). The subsequent increase or decrease in the value of $\Delta C$ from each of these re-runs quantifies the dominance of $\simeq 3.5$ keV photons in each individual spectrum. We find that the jackknifed iterations in bins 1, 3 and 4 do not result in a significance change in the $\Delta C$ values in the region of interest. However, in bin 2, where there is evidence for a $\simeq 3.5$ keV excess in the joint fit when all 29 clusters are included, we find a significant variation in $\Delta C$ during the jackknifing (Fig.~\ref{fig:jackknifedplot}).  This strongly implies that the detection of a $\simeq 3.5$ keV excess in Figure~\ref{fig:tx-all-bin-fits-mc}(b) is being driven by a subset of the clusters in the bin and is not a global feature.

\subsection{Joint fits to the full sample}
\label{subsec:resultsAll}

\begin{figure}
    \includegraphics[width=0.5\textwidth]{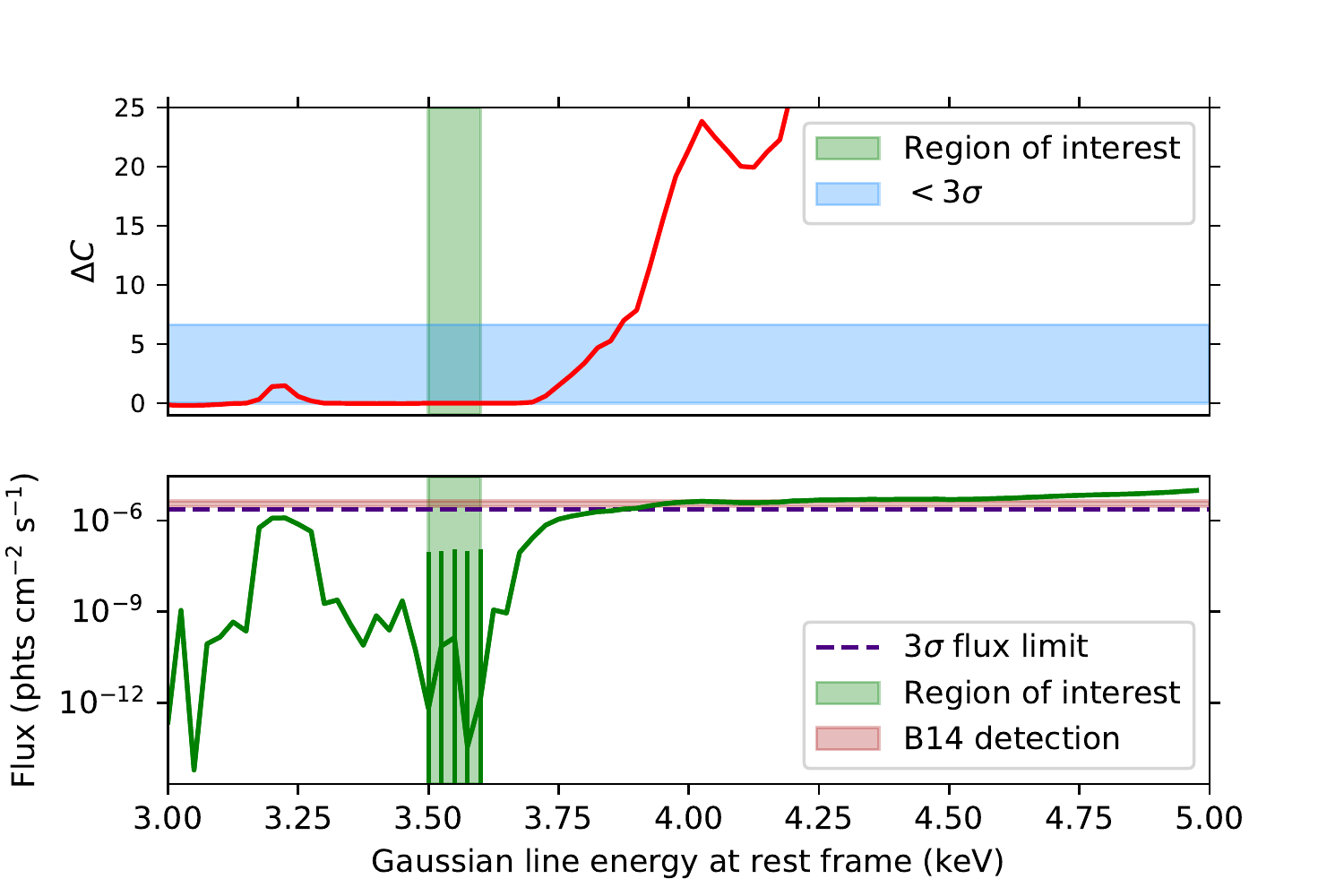}
    \caption{Similar to Figure~\ref{fig:tx-all-bin-fits} showing the trend in $\Delta C$ using 114 clusters in the sample, minus the three clusters with known flux excess at $\simeq 3.5$ keV -- XCS J0003.3+0204, XCS J1416.7+2315 and XCS J2223.0-0137. In the bottom panel, the pink horizontal shaded region shows the constraints from B14 for 73 clusters (using PN data only). The dashed purple line corresponds to the $3\sigma$ flux limit defined for the sample. The fitted abundance for this analysis was $Z=0.24 Z_{\odot}$.}
    \label{fig:allclustersmc}
\end{figure}

To obtain the highest possible sensitivity to a spectral feature arising from dark matter, we have carried out joint analysis using all 114 clusters without an individual $\simeq 3.5$ keV excess. In this case, flux errors have only been calculated in the region of interest due to the excessive computation required for the $3 < E < 5$~keV energy range. The results are presented in Figure~\ref{fig:allclustersmc}. No significant improvement in the fit is found in the region of interest. We measure a best fit Gaussian flux of $1.40 \times 10^{-10}$ photons cm$^{-2}$ s$^{-1}$ at 3.55 keV, corresponding to a mixing angle $\sin^{2}2\theta = 2.5 \times 10^{-15}$. We note that during the fit, {\sc xspec} failed to compute a lower limit on this value, due to the lack of flux excess in the region of interest. 

To demonstrate that this lack of evidence is not a reflection of a lack of sensitivity, we have included on the lower panel of Figure~\ref{fig:allclustersmc} an estimate of the $3\sigma$ upper limit on the flux (dashed purple line) of $F_{\textrm{DM}} = 2.41 \times 10^{-6}$ photons cm$^{-2}$ s$^{-1}$. The $3\sigma$ upper limit corresponds to the measured flux where the fit improvement (red line) is equivalent to the $3\sigma$ threshold for a detection (blue band). We assume negligible impact from the {\tt ARF} on the flux limit across the specified energy range. 

\begin{figure}
    \includegraphics[width=0.46\textwidth]{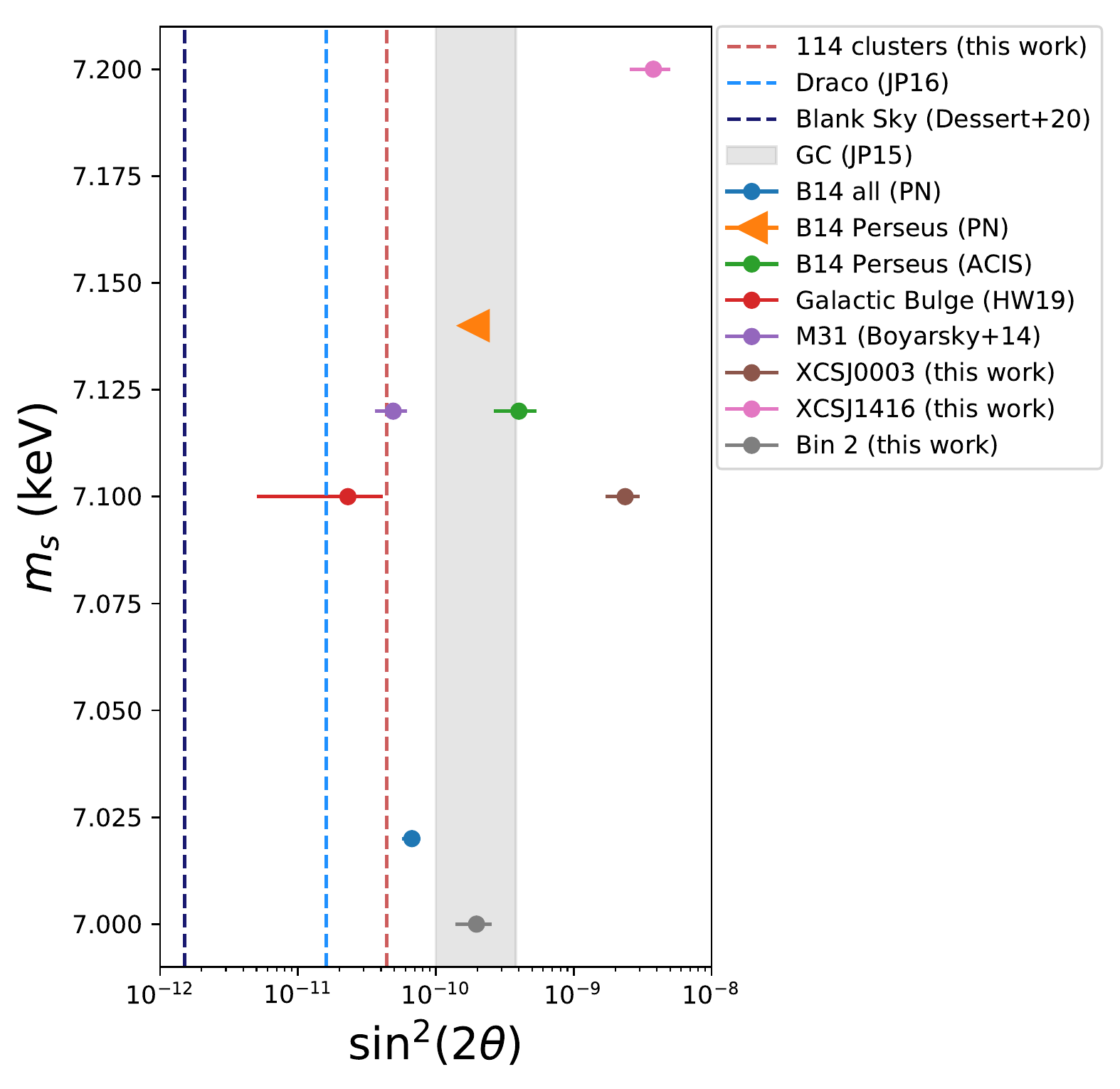}
    \caption{Comparison of the constraints on the sterile neutrino mixing angle and mass from various studies in the literature. The constraint on the mixing angle from the 114 clusters in this work is given by the pink dashed vertical line. Blue and purple dashed lines indicate upper limits from non-detections in Draco and blank sky observations, respectively. The grey shaded region highlights the range of mixing angles from the detection of a line in the Galactic Centre, depending on the choice of dark matter profile used. The remaining points correspond to detections of the line in various astrophysical systems.}
    \label{fig:mixingangleslit}
\end{figure}

The 114 clusters in the joint fit have a weighted mass per distance squared of $1.65 \times 10^{10} M_\odot$ Mpc$^{-2}$, which corresponds to a maximum mixing angle, $\sin^{2}2\theta = 4.4 \times 10^{-11}$. This is the most stringent mixing angle constraint obtained from our analysis - it is well below the values in Table~\ref{tab:3.5kevprops} for individual clusters and bin 2 (with all 29 clusters included). Comparisons with the B14 analysis are indicated in the bottom panel of Figure~\ref{fig:allclustersmc}, where the the red shaded region highlights the flux estimate obtained on the $\simeq$ 3.5~keV line using the stacked PN spectrum of 73 clusters. As can be seen in the plot, the upper limit of the flux as a function of energy (given by the dashed purple line) is $2\sigma$ below the preferred B14 value for the line using {\em XMM}-PN data of 73 clusters ($\sin^2(2\theta)=6.7_{-1.7}^{+2.7} \times 10^{-11}$). A summary of the mixing angles obtained from individual and jointly fitted clusters in this study, alongside constraints obtained from the literature (note that this is an incomplete list), is displayed in Figure \ref{fig:mixingangleslit}.
 
\section{Discussion}\label{sec:discussion}

It is clear from Figures~\ref{fig:tx-all-bin-fits}, \ref{fig:tx-all-bin-fits-mc}, \ref{fig:jackknifedplot}, and \ref{fig:allclustersmc} that a $\simeq 3.5$ keV flux excess is not a ubiquitous feature in cluster spectra. As Figure \ref{fig:tx-all-bin-fits-mc}(b) shows, where a flux excess exists, its strength does not increase with cluster temperature (and hence halo mass). Therefore, it seems unlikely that these and previously reported `3.5 keV line' detections have a dark matter origin. In this section we investigate possible reasons why three clusters show an excess of emission at $\simeq$ 3.5 keV (Sect.~\ref{sec:clustercontributions}), and test the robustness of our analysis methods, to ensure that these are not somehow artificially masking a feature related to dark matter decay (Sect.~\ref{subsec:systematics}).

\subsection{Individual clusters with excess emission at $\simeq 3.5$ keV}
\label{sec:clustercontributions} 

\subsubsection{XCS J0003.3+0204}
\label{subsec:xcs0003}

\begin{figure}
    \includegraphics[width=0.45\textwidth]{29-fit.pdf}
    \includegraphics[width=0.45\textwidth]{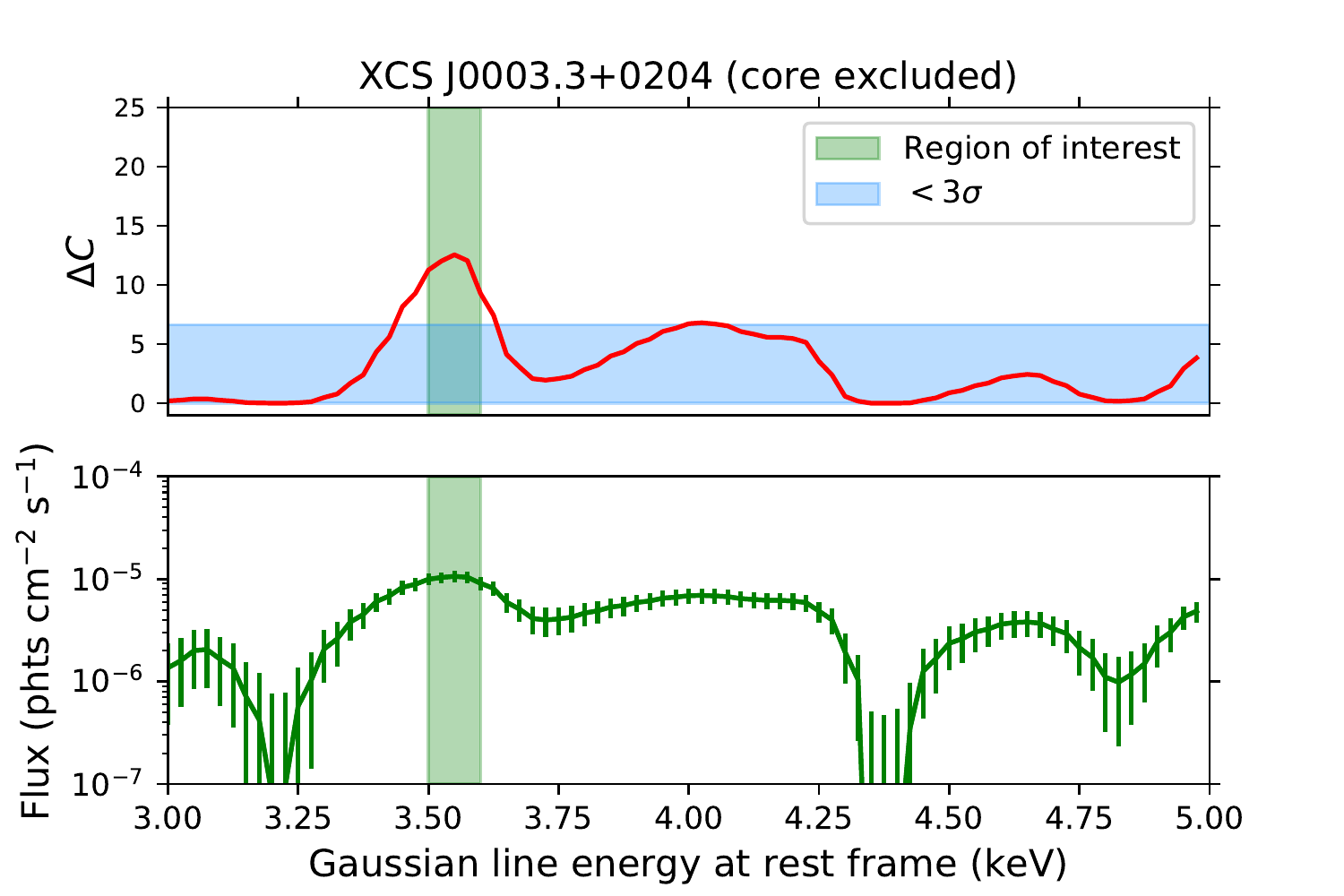}
    \caption{Plots showing the trend in $\Delta C$ (see Fig.~\ref{fig:tx-all-bin-fits} for full description) for the cluster XCS J0003.3+0204.  The top plot displays the analysis using a spectrum with the core-included (i.e. our standard analysis) and the bottom plot shows the trend using a spectrum with the core region excluded (see Sect.~\ref{subsec:xcs0003}).}
    \label{fig:RXCtests}
\end{figure}

The cluster XCS J0003.3+0204 \citep[better known as Abell 2700,][]{Abell2440ref}, and first identified in X-rays by ROSAT \citep[][]{ROSAT} has a  RM ID = 2789, a RM redshift of $z_{\rm phot}^{\rm RM}=0.11$ and a RM richness of $\lambda=38.9$. This well-studied cluster is not reported as having AGN activity or any distinct morphological or galaxy properties \citep[e.g.][]{Ettori2015,Lovisari2018, Bohringer2007,Bohringer2010b, Holland2015}. 
The best fit temperature and metallicity (following method A, see Sect.~\ref{sec:sample}) are $T_{\rm X}^{\rm PN}=4.78_{-0.12}^{+0.12}$ keV and $Z=0.4_{-0.04}^{+0.04}Z_\odot$ respectively. The fit values quoted here are based on \textit{XMM}-PN observation ObsID {\tt 0201900101} (Fig.~\ref{fig:RXCimage}). This observation was made on 2004-06-24, and has a flare corrected exposure time of 19 ks. We note that the rate of flaring in the raw events file is less than 2\% for this observation. There are no other \textit{XMM}-PN observations available for this cluster, so we cannot investigate any possible variability in the $\simeq 3.5$ keV excess for this cluster. 
A comparison of the $\Delta C$ analysis between the original spectrum and one where the core region $r<0.15R_{500}$ is excluded, is shown in Figure \ref{fig:RXCtests}. We find that the shape of the $\simeq$ 3.5 keV excess is largely insensitive to the removal of the $r<0.15R_{500}$ region. Finally, we check all available MOS data for XCS J0003.3+0204 for evidence for a 3.5 keV feature. Given the MOS camera is approximately half as sensitive as the PN, we do not expect to detect a feature at the same significance. The comparison of PN and MOS data for this cluster is shown in Figure \ref{fig:moscomparisons}(a). We observe a feature of similar shape in the MOS2 data at a slightly higher energy ($\simeq 3.6$ keV), however, there is no clear evidence of a feature within the region of interest.     
 
\begin{figure*} 
    \begin{centering}
    \includegraphics[trim={1.5cm 1.5cm 1.5cm 1.5cm},clip, width=0.8\textwidth]{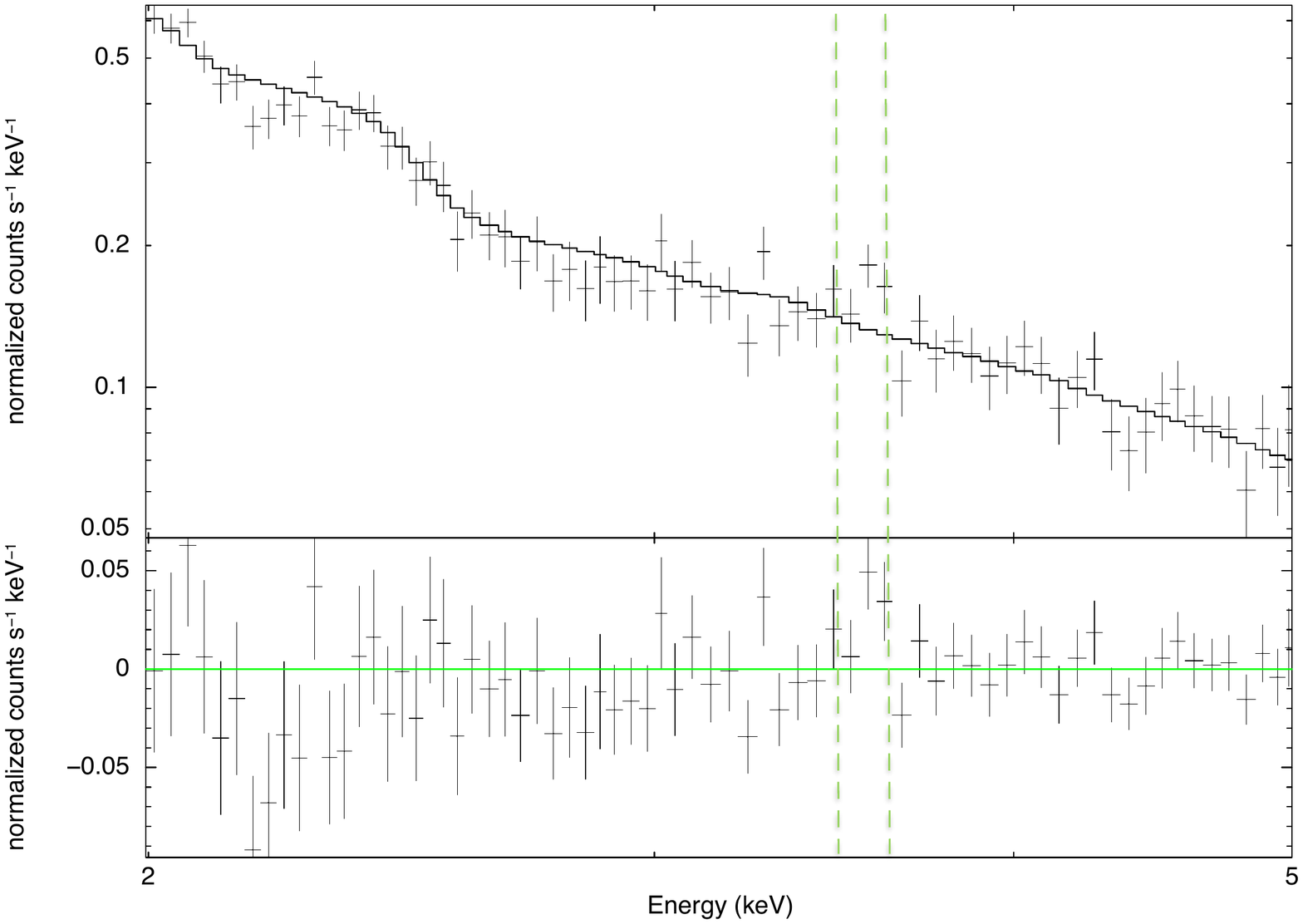}
    \caption{Spectrum of the cluster XCS J0003.3+0204 (located in ObsID {\tt 0201900101}) fitted using model A outlined in Section~\ref{subsec:jointfitting}. The top panel shows the spectrum and fitted model across the $2 - 5$ keV energy range. The bottom panel shows the the residuals i.e. the difference between the model and the spectrum. For visual purposes, the spectrum has been grouped such that each bin has a SNR $\geq 15$.  The dashed green lines enclose the $3.5-3.6$ keV region of interest.}
    \label{fig:RXCspectrum}
    \end{centering}
\end{figure*}

From the existing analysis/data, it is not possible to unambiguously explain the flux enhancement at $\simeq 3.5$ keV in XCS J0003.3+0204. It is unlikely that the enhancement is related to background flare contamination, since ObsID {\tt 0201900101} is one the cleanest of the sample (of 117). The symmetrical shape of the $\Delta C$ feature, and the fact that an enhancement is visible in the spectrum (Fig.~\ref{fig:RXCspectrum}), supports the interpretation that this is a genuine astrophysical emission feature, either from the cluster itself, or from a system along the line of sight. Despite this feature being stronger than would be expected from dark matter decay, it is not obvious that the excess is the result of a plasma transition or charge exchange process, since such a feature would also be present in other systems. One interpretation for this excess could lie in the treatment of point sources in the spectrum of XCS J0003.3+0204. As shown in Figure \ref{fig:RXCimage}, all point sources have been masked from the cluster observation. However, it is possible that some excess point source emission is persisting within the source region, which could be influencing the modelling of the continuum within the $\simeq 3.5$ keV range (impacting the overall shape of the background/source spectrum).

This cluster is the only one of the 117 that displays a conclusive 3.5 keV feature at the $>3\sigma$ level, so it is rare. Specifically, the detection of a 3.5 keV feature in XCS J0003.3+0204 constitutes only the second ever detection of a line in a single cluster (the first being Perseus). To examine just how rare, we plan to apply our $\Delta C$ technique to the other 228 (346-117) clusters with measured $T_{\rm X}$ values in the G20 sample.

\subsubsection{XCS J1416.7+2315}
\label{subsec:xcs1416}

The cluster XCS J1416.7+2315 \citep[first detected in X-rays by ROSAT, e.g.][]{ROSAT}, and also known as RX J1416.4+2315 \citep{KathyPaper} has a RM ID=5527, a RM redshift of $z_{\rm phot}^{\rm RM}=0.137$ and a RM richness of $\lambda=31.7$. Based upon the best fit parameters to the \textit{XMM}-PN spectrum using model A, described in Section \ref{sec:sample}, the cluster has a measured temperature and metallicity of $T_{\rm X}^{\rm PN}=3.28_{-0.12}^{+0.12}$ keV and $Z=0.17_{-0.05}^{+0.05}Z\odot$ respectively. It is noteworthy that this system has a comparatively low metal abundance compared to the average obtained in bin 1 (see column 5 of Table~\ref{tab:props}). The $\simeq$ 3.5 keV excess, i.e. the region where the $\Delta C$ value is $>3\sigma$ is significantly wider than the spectral energy resolution of the \textit{XMM}-PN detector ($\Delta E=88$ eV). 

\begin{figure}
    \begin{centering}
    \includegraphics[width=0.45\textwidth]{95-fit.pdf}
    \includegraphics[width=0.45\textwidth]{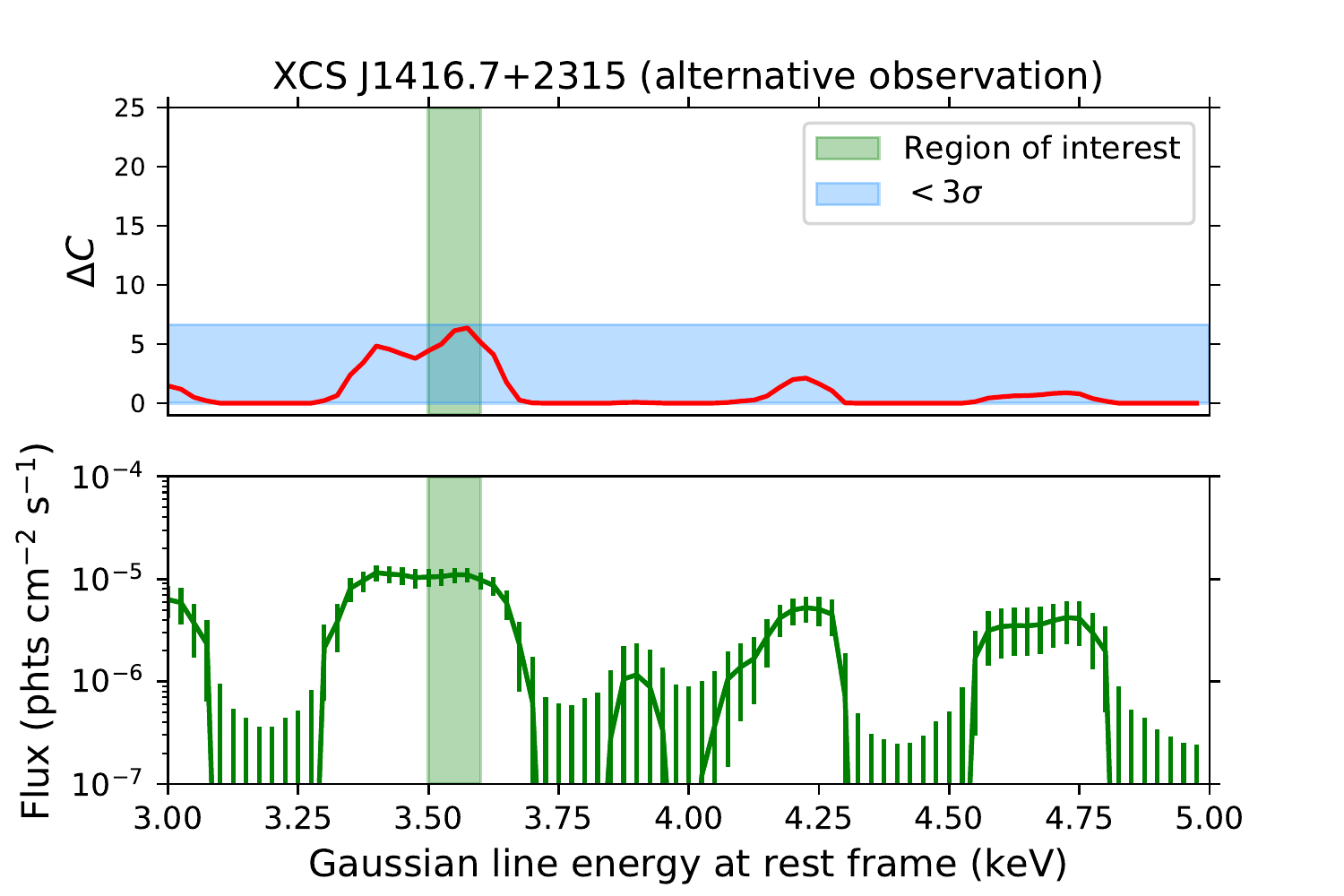}
    \includegraphics[width=0.45\textwidth]{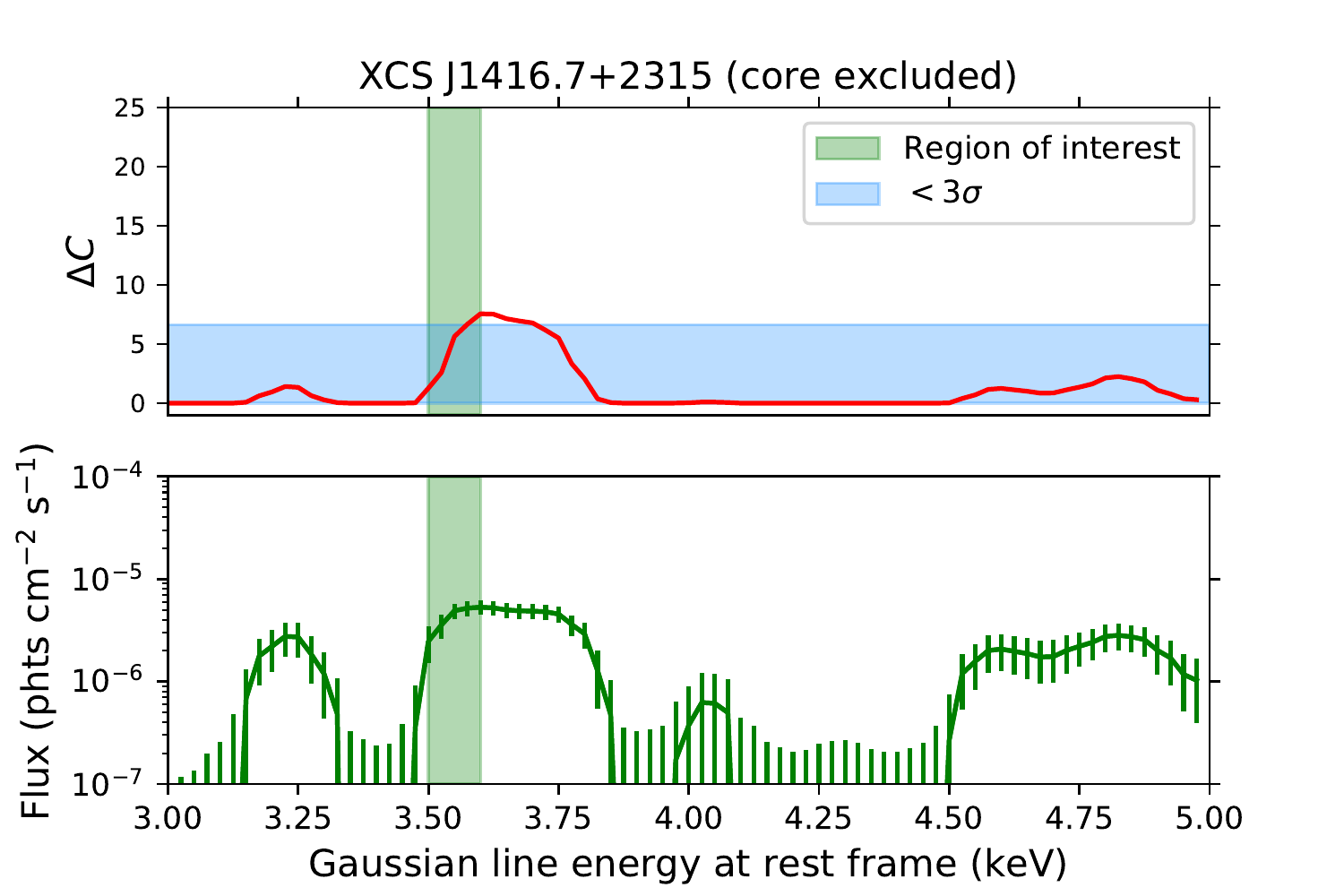}
    \caption{Plots showing the trend in $\Delta C$ (see Fig.~\ref{fig:individualclusters} for full description) for the cluster XCS J1416.7+2315.  The top plot shows the analysis using the {\em XMM} ObsID 0722140401 (i.e. our standard analysis), the middle plot shows the analysis performed using the {\em XMM} ObsID 0722140101 (see Sect.~\ref{subsec:xcs1416}) and the bottom plot shows the analysis performed using ObsID 0722140401 with the core region excluded (see Sect.~\ref{subsec:xcs1416}).}
    \label{fig:RXJtests}
    \end{centering}
\end{figure}

The analysis presented in Section \ref{subsec:resultsIndiv}, and the best fit temperature and abundance quoted above, are based on the {\em XMM} observation with ObsID {\tt 0722140401}. This observation was taken on 2014-01-31, and has a cleaned exposure time of 18 ks. However, this cluster has been the target of another \textit{XMM} observation ({\tt 0722140101}). This observation was made on 2014-01-03, and has a flare corrected exposure time of 4 ks. The availability of two observations of the same cluster, made roughly a month apart, give us the opportunity to look for time variability in the excess flux at $\simeq$ 3.5 keV. A comparison of the analysis between the two observations (using PN data only) is shown in Figure \ref{fig:RXJtests} (top vs middle). The shorter observation (middle panel) shows a noticeably different shape of the $\Delta C$ excess at $\simeq$ 3.5 keV, and a drop in the maximum value of $\Delta C$ in the region of interest to below $3\sigma$. There are several possible causes for these differences. For example, they could be due to poor photon statistics in the shorter observation. Alternatively, they could be due to the differing effects of background flaring, which rises from 35\% of the raw events list for ObsID {\tt 0722140401} to 90\% for ObsID {\tt 0722140101}. Assuming the measurement of a shape and flux change is robust, then the most likely astrophysical interpretation for a time-dependent signal would be AGN variability: XCS J1416.7+2315 is described in the literature as a fossil cluster with known variable AGN activity, e.g. \cite{Miraghaei2015}.
 
We have also investigated whether the presence of an excess at $\simeq$ 3.5 keV in XCS J1416.7+2315 might be associated with a cool-core. A comparison of the $\Delta C$ analysis between the original spectrum and one where the inner $0.15R_{500}$ is excluded is shown in Figure \ref{fig:RXJtests} (top vs bottom). The removal of photons from the cluster core does not significantly change the shape of the $\Delta C$ excess at $\simeq$ 3.5 keV. The significance of the enhancement is lower after core removal, but remains in excess of $3\sigma$ in the region of interest.

We repeat our analysis on this cluster using available MOS data, shown in Figure \ref{fig:moscomparisons}(b). Yet again, owing to the differing sensitivities of the PN and MOS cameras, we do not expect to recover a significant feature in the MOS data. Interestingly, we note two narrower features in the MOS data which align broadly with the energy of the $\simeq 3.5$ keV excess in the PN spectrum of XCS J1416.7+2315. 

Again, from the existing analysis/data it is not possible to unambiguously explain the flux enhancement at $\simeq 3.5$ keV in XCS J1416.7+2315. However, its broad and asymmetrical shape is not consistent with a discrete emission line origin. An additional {\em XMM} observation would be needed to explore the hint of time dependence seen in Figure~\ref{fig:RXJtests} (top vs middle). If confirmed, then AGN activity would be the most likely cause of the variability (and potentially of the $\simeq 3.5$ keV flux excess). In that case, follow-up with \textit{Chandra} would assist with resolving the central point source. Both of the {\em XMM} observations of this cluster ({\tt 0722140401} and {\tt 0722140101}) were taken during times of enhanced background flaring (especially {\tt 0722140101}). It would be possible to explore the impact of background flaring on the goodness of fit of model A and model B at $\simeq 3.5$ keV, by relaxing/tightening the criteria used to reject time periods affected by flares in these two observations.

\subsubsection{XCS J2223.0-0137}
\label{subsec:xcs2223}

The cluster XCS J2223.0-0137 \citep[better known as Abell 2440,][]{Abell2440ref}, and first identified in X-rays by HEAO-1 \citep[][]{HEAO1} has a RM ID = 48, a RM redshift of $z_{\rm phot}^{\rm RM}=0.101$ and a RM richness of $\lambda=90.7$. The best fit temperature and metallicity (following method A, see Sect.~\ref{sec:sample}) are $T_{\rm X}^{\rm PN}=4.39_{-0.10}^{+0.08}$ keV and $Z=0.4_{-0.04}^{+0.04}Z\odot$ respectively. The fit values quoted here are based on \textit{XMM}-PN observation ObsID {\tt 0401920101}. This observation was made on 2006-11-18, and has a cleaned exposure time of 23 ks, and a background flaring rate of 35\%. Due to the fact that this cluster only has one available {\em XMM} observation, a variability analysis cannot be performed. Furthermore, we forgo an analysis excluding the central regions of the cluster because XCS J2223.0-0137 is a complex merging system \citep[see e.g.][]{Abell2440Mohr,Abell2440Maurogordato} with two distinct peaks in the X-ray emission, making the exclusion of the cluster core problematic. We do, however, study the available MOS data for this cluster (Fig. \ref{fig:moscomparisons}(c)), finding no clear evidence of an excess at $\simeq 3.5$ keV. 

We argue that the broad ($3.25<E<3.85$ keV) and multi-peaked shape of the $> 3 \sigma$ flux excess shown in Figure~\ref{fig:individualclusters} (bottom) is not consistent with being associated with a discrete emission line. Due to the complex cluster morphology, we forgo further discussion into the nature of the behaviour of the $\Delta C$ of XCS J2223.0-0137 in the range $3.25<E<3.85$ keV. 

\subsection{Methodology validation}
\label{sec:methodvalid}
\label{subsec:systematics}

In this section, we investigate the influence of various aspects of our methodology on the results presented herein: the blueshifting technique (Sect.~\ref{subsec:blueshiftingtests}), alternative weighting methods (Sect.~\ref{subsec:plasmaweighting}), solar abundance tables (Sect.~\ref{subsec:solarabundances}), photoelectric absorption models (Sect.~\ref{subsec:nhvalue}), the use of photometric redshifts (Sect.~\ref{subsec:photo-z}), and the choice of plasma code ({Sect.~\ref{subsec:plasma-code}}).

\subsubsection{Blueshifting}
\label{subsec:blueshiftingtests}

\begin{figure}
    \begin{centering}
    \includegraphics[width=0.45\textwidth]{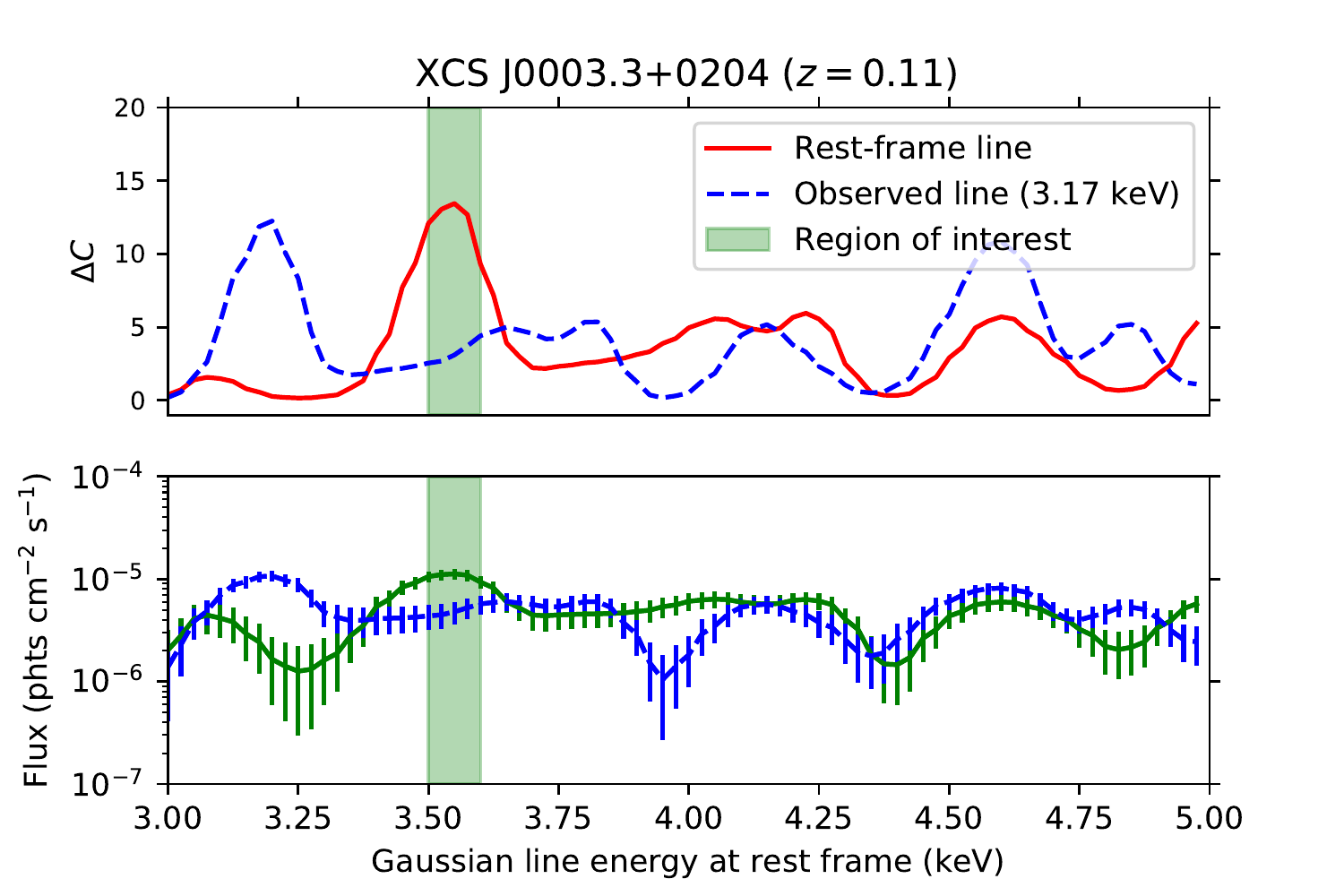}
    \caption{Plot of the change in fit statistic, $\Delta C$ (as in Figure~\ref{fig:tx-all-bin-fits}) for the cluster XCS J0003.3+0204.  Our standard analysis is given by the red solid line and the blue dashed line shows the change in $\Delta C$ with the analysis perform in the observed frame, i.e. without the blueshifting step (see Sect.~\ref{subsec:blueshiftingtests}).}
    \label{fig:blueshiftcomps}
    \end{centering}
\end{figure}

We test our blueshifting technique using cluster XCS J0003.3+0204 (see Fig. \ref{fig:blueshiftcomps}). For this, we repeat the fit to model B without carrying out the blueshifting step. As expected, we find that the flux excess at $\simeq$ 3.5 keV now appears at the observed rather than rest frame energy, i.e. at the expected value ($\sim 3.2$ keV) for a $z_{\rm phot}^{\rm RM}=0.11$ system (see blue dashed line in Fig.~\ref{fig:blueshiftcomps}). 

\subsubsection{Flux weighting}
\label{subsec:plasmaweighting}

The weighting technique described in Section~\ref{sec:weighting} includes the implicit assumption that any excess flux at $\simeq 3.5$ keV is due to dark matter decay. However, if the flux at that energy was instead a result of emission from the ICM, then the use of a mass-dependent weighting would be inappropriate. Therefore, we test an alternative method of weighting based only on the cluster redshift,
\begin{equation}\label{eq:plasmaflux}
    w'_{i} = \frac{1+z_i}{4\pi d^2_{i,L}},
\end{equation}
and rerun the joint fits in each temperature bin, finding almost identical results to Figure \ref{fig:tx-all-bin-fits} (i.e. no $>3 \sigma$ detection of a flux excess in any of the bins).

\subsubsection{Solar abundance tables}\label{subsec:solarabundances}
\label{subsec:plasma-code}

Our default method uses the \cite{AndersandGrevesse} solar abundance table. We have also run the joint fits in each temperature bin using the \cite{AsplundGrevesse}, \cite{Lodders} and \cite{GrevesseSauval} abundance tables, recovering almost identical results to Figure \ref{fig:tx-all-bin-fits}, (i.e. no $>3 \sigma$ detection of a flux excess in any of the bins). Nevertheless, we note the best fit metal abundance for each $T_{\rm X}$ bin does increase slightly compared to the values in column 6 of Table~\ref{tab:props}.

\subsubsection{Photoelectric absorption}
\label{subsec:nhvalue}

Our default method uses the {\tt tbabs} implementation of photoelectric absorption in {\sc xspec} because the {\tt wabs} model is now considered to be outdated \citep{WilmsWabs}. However, the {\tt wabs} model was implemented in other previous studies of the $\simeq 3.5$ keV flux excess, including B14, so we have also run the joint fits in each temperature bin using {\tt wabs} for comparison. Once again, we find almost identical results to Figure \ref{fig:tx-all-bin-fits}, (i.e. no $>3 \sigma$ detection of a flux excess in any of the bins).

\subsubsection{Use of photometric redshift measurements}
\label{subsec:photo-z}

The ensemble behaviour of the $z_{\textrm{phot}}^{\rm RM}$ is well understood. According to \cite{RM} the scatter in the photometric redshift measurements is $\sigma_z \approx 0.006$ at $z \approx 0.1$, increasing to $\sigma_z \approx 0.020$ at $z \approx 0.5$. The median value for $|\Delta z|/(1+z)$ for the full sample is 0.006, where $\Delta z = z_{\textrm{phot}} - z_{\textrm{spec}}$. Therefore, the $1\sigma$ error in energy in blueshifting a 3.55 keV line to the local frame ranges from 20 eV for a cluster at $z \approx 0.1$ to 30 eV at $z \approx 0.5$, which is well below the spectral energy resolution of the PN detector ($\Delta E=88$ eV). Therefore, it is unlikely that the use of $z_{\textrm{phot}}^{\rm RM}$ values is the reason for a non detection of a $>3 \sigma$ detection of a $\simeq$ 3.5 keV flux excess in our binned analyses shown in Figures~\ref{fig:tx-all-bin-fits} and \ref{fig:allclustersmc}. For our joint fits, we conclude that the ensemble scatter in $z$ is applicable.  

However, errors in estimates of $z_{\textrm{phot}}^{\rm RM}$ for individual clusters may influence the results discussed in Section~\ref{subsec:resultsIndiv}, if they exceed the ensemble average. For the three individual clusters described in Section~\ref{sec:clustercontributions}, the spectroscopic redshift for XCS J1416.7+2315 \citep{Romer2001} is almost exactly the same as the RM value, $z_{\textrm{phot}}^{\rm RM}=0.137$. However, according to \cite{StrubleRood}, XCS J0003.3+0204 (or Abell 2700) has $z_{\textrm{spec}}= 0.0924$ based on 9 cluster members, and XCS J2223.0-0137 (or Abell 2440) has $z_{\textrm{spec}}= 0.0906$ based on 48 cluster members (compared to $z_{\textrm{phot}}^{\rm RM}=0.11$ and $=0.101$ respectively).  Therefore, we have refitted the spectrum of XCS J0003.3+0204 and XCS J2223.0-0137 using the spectroscopic values. The results are shown in Figure~\ref{fig:spectrozs}. We conclude that there is negligible impact from using spectroscopic redshifts, since the $\simeq 3.5$ keV excess in both clusters remains within the region of interest. 

\subsubsection{Choice of plasma code}

Searches for new emission lines are sensitive to both the temperature and metal abundance. Therefore it is important that these properties are measured precisely to prevent erroneous detections (or non-detections) of excess flux at $\simeq 3.5$ keV. As shown in detail in \cite{mernier2019atomic} the two codes used most in the field of X-ray cluster spectroscopy, {\sc AtomDB} and {\sc Spexact}, do not produce consistent results for metal abundance for low temperatures plasmas. For $T_{\rm X} \leq 2$ keV, the discrepancies can be up to 20$\%$ in the Fe abundance. The {\sc Spexact} code is not implemented inside {\sc xspec}\footnote{{\sc Spexact} is implemented in the SPEX fitting package, {\tt www.sron.nl/astrophysics-spex}.}, so it is not possible to conduct a direct comparison here. However, only 7 clusters in our sample have measured $T_{\textrm{X}}$ values below 2 keV (and all in bin 1). Even if this issue impacts the results in plot (a) of Figure~\ref{fig:tx-all-bin-fits}, it will not impact the results shown in other three plots.

\section{Summary and Conclusions}
\label{sec:conclusions}

In this work, we have used a similar technique to the seminal \cite{Bulbul2014}, B14, paper, in order to explore the evidence for an $\simeq$ 3.5~keV flux excess in the spectra of clusters of galaxies.  We used individual and joint fits to \textit{XMM}-PN spectra of 117 SDSS redMaPPer galaxy clusters ($0.1 < z < 0.6$, 1.7 < $T_{\rm X}$ < $10.6$ keV). This is the largest study of its kind to date. By comparison, the B14 study used a sample of 73 clusters.

The analysis of the individual spectra identified three systems with an excess of flux at $\simeq$ 3.5~keV. This refers to an excess over the fiducial plasma model, taking into account one additional degree of freedom. None of these individual clusters are the most dark-matter dominated or nearest systems in our study (see Table \ref{tab:data}). In two of the three cases (XCS J1416.7+2315 and XCS J2223.0-0137), the flux excess, as a function of energy, is not consistent with a discrete emission feature due to the feature's broad, asymmetrical shape in both cases.

In the remaining case (XCS J0003.3+0204), the excess may result from a discrete emission line with a central energy of $E=3.55$ keV. This feature, however, is unlikely to have a dark matter origin for two reasons. First, this is the only cluster in the sample to show such a feature, and yet there are many other observations of similar or better sensitivity in the sample. Second, the estimated flux ($11.2_{-0.31}^{+0.31} \times 10^{-6}$ photons cm$^{-2}$ s$^{-1}$) results in significantly higher mixing angle constraints ($\sin^2(2\theta)=2.36_{-0.65}^{+0.65} \times 10^{-9}$) than the ones presented from the stacked analysis of 73 \textit{XMM}-PN spectra in B14. The line strength for XCS J0003.3+0204 is most comparable to the {\em XMM}-MOS detection for the Perseus cluster (core-included) in B14. Nevertheless, there exists an order of magnitude of difference in the derived mixing angles ($\sin^2(2\theta) \simeq 6 \times 10^{-10}$, B14). 

We note that this is only the second time that a significant detection of a $\simeq$ 3.5~keV line-like excess has been measured in an individual cluster (the other being in Perseus). Furthermore, unlike Perseus, the strength of the 3.5~keV excess in XCS J0003.3+0204 shows almost no dependence on the removal of the core region from the spectrum. Flaring is also unlikely to be causing such an excess as we report a very low flare rate (less than 2\%) for this observation. 

The primary motivation for our study was a search for evidence of an increase in the $\simeq$ 3.5~keV flux excess with $T_{\textrm{X}}$. Such evidence would firmly support the dark matter interpretation (and vice versa if the excess weakens with $T_{\textrm{X}}$) because $T_{\textrm{X}}$ is a reliable tracer of the underlying halo mass. A temperature-dependent search would additionally eliminate the possibility of a plasma line masking an emission line of dark matter origin as the relevant plasma lines in the region of interest weaken with temperature (contrary to dark matter).

We therefore grouped the remaining 114 clusters into 4 roughly equally sized $T_{\textrm{X}}$ bins, and performed joint fits in each bin. We did not find evidence of a significant excess in flux at $\simeq$ 3.5~keV in any of the bins. Therefore, from our study, we cannot comment on whether (if it exists at all) the $\simeq$ 3.5~keV flux excess gets stronger or weaker with mass. Repeating the joint fits in each bin with the inclusion of the three clusters with excess emission resulted in a significant joint detection in one bin (bin 2). However, after performing a jackknife analysis on the clusters in this bin, it is found that the joint detection is dependent on the $\simeq$ 3.5~keV flux excess in two individual clusters (XCS J0003.3+0204 and XCS J2223.0-0137).

We maximised sensitivity to a potential weak dark matter decay feature at $\simeq$ 3.5~keV, by performing a joint fit across all 114 clusters. Again, no significant excess was found at $\simeq$ 3.5~keV. A best fit Gaussian flux of $1.40 \times 10^{-10}$ photons cm$^{-2}$ s$^{-1}$ was measured at 3.55 keV, corresponding to a mixing angle $\sin^{2}2\theta = 2.5 \times 10^{-15}$ (note that errors are not quoted due to the inability of {\sc xspec} to compute formal uncertainties on a small flux within the region of interest).

Furthermore, we estimated a $3\sigma$ upper limit of an undetected emission line at $\simeq$ 3.5~keV to be $F_{\textrm{DM}} = 2.41 \times 10^{-6}$ photons cm$^{-2}$ s$^{-1}$. The resulting maximum mixing angle from our 114 clusters is then $\sin^2(2\theta)=4.4 \times 10^{-11}$, lower than the previous estimates for favoured mixing angles from cluster studies. These include the {\em XMM}-PN value for 73 clusters in B14 ($\sin^2(2\theta) \simeq 7 \times 10^{-11}$) and \cite{Bulbul2016} study of {\em Suzaku} observations for 47 clusters ($\sin^2(2\theta) \simeq 6 \times 10^{-11}$). Moreover, our result is comparable to among the most stringent constraints on the non-detection of a dark matter decay feature using {\em XMM} observations of Draco \cite[$\sin^2(2\theta) \simeq 2 \times 10^{-11}$,][]{DracoJeltemaandProfumo} and {\em XMM} blank sky observations \citep{DessertBlankSky2018}. Relevant constraints from previous studies are summarised in Figure \ref{fig:mixingangleslit}.

We conclude that although there is a measurable flux excess at $\simeq$ 3.5 keV in some cluster spectra (e.g. XCS J0003.3+0204), this is not a ubiquitous feature, and hence unlikely to originate from sterile neutrino dark matter decay. We have carried out a series of checks to demonstrate that our methodology is not artificially masking the existence of a weak dark matter decay feature. We perform tests on our blueshifting technique to ensure the purported 3.5 keV feature appears at the correct rest-frame energy, in addition to alternative weighting methods, abundance and photoelectric absorption models, and comparisons with spectroscopic data. We have found our methodology to be robust to these tests and have negligible impact on our presented results. 

Future work on the constraining the origin of the purported 3.5 keV feature will be informed most directly by the launch of the {\em XRISM} satellite (the successor to the short-lived {\em Hitomi} mission). With significantly improved spectral resolution, it will be possible to determine the precise energies of elemental and unknown lines, to clarify whether a $\simeq$ 3.5 keV flux excess is indeed originating from a discrete emission line (such as in the case of XCS J0003.3+0204). Moreover, deeper observations of single clusters with claimed $\simeq$ 3.5 keV emission such as Perseus will be able to confirm whether such a line exists, and if so, to what extent it is resolvable from the nearby K and Ar transition lines.   

We further aim to revisit previous analyses such as B14, in which the cluster sample is produced using publicly available {\em XMM} observations. Given the complications associated with stacking methods and the fact that individual systems can contribute significantly to the appearance of a line in joint cluster searches, we will repeat the analysis in B14 simply by jointly fitting all available clusters in parallel, in addition to fitting each cluster individually. The aim of this would be to see if a $\simeq$ 3.5 keV excess is detected in any individual clusters in the B14 analysis, which could suggest these clusters are responsible for an overall so-called dark matter decay feature (or masking one). 

Given that two of the three clusters in this study with a measured flux excess at $\simeq$ 3.5 keV also display high rates of flaring, investigating the rate of flaring across all clusters which might contain such flux excesses would be a useful diagnostic to examine whether the origin of the line is instrumental. Deeper, repeated observations of individual clusters are also needed to further test the possibility of a variable $\simeq$ 3.5 keV feature (e.g. there is a hint of variability in the feature observed in XCS J1416.7+2315), which could lend support to the interpretation that $\simeq$ 3.5 keV emission arises from AGN variability (or the interaction between AGN and ALPs in a more exotic dark matter scenario).  

Finally, further work to conclusively determine the existence and properties of the intriguing 3.5 keV feature will require even larger jointly fitted cluster samples. Hence, future work on this topic will involve a repeat of this analysis on a larger sample of redMaPPer selected clusters in the Dark Energy Survey Year 3 footprint with associated archival X-ray data. 

\section*{Acknowledgements}

The authors thank K. Abazajian for helpful discussions. SB is grateful to F. Hofmann for advice on X-ray analysis, S. Profumo for comments, and A. R. McDaniel for guidance with calculations. The authors would also like to thank the referee for useful suggestions. SB, KR, and PG acknowledge support from the UK Science and Technology Facilities Council via grants ST/N504452/1 (SB), ST/P000525/1 (KR, PG). TJ was supported by the National Science Foundation under Grant No. 1517545. MH acknowledges financial support from the National Research Foundation and the University of KwaZulu-Natal. PTPV was supported by Funda\c{c}\~{a}o para a Ci\^{e}ncia e a Tecnologia (FCT) through research grants UIDB/04434/2020 and UIDP/04434/2020. This research made use of {\sc Astropy}\footnote{http://www.astropy.org}, a community-developed core Python package for astronomy \citep{astropy:2013, astropy:2018}, as well as {\sc numpy}, {\sc scipy} and {\sc matplotlib}. The data underlying this study are available in the article.




\bibliographystyle{mnras}
\bibliography{SIDM} 




\appendix

\section{Full data sample}

Table~\ref{tab:data} provides the X-ray properties and associated projected dark matter masses for the 117 clusters used in this study.

\begin{table*}
\begin{center}
\caption[]{{\small Properties of the cluster sample. XCSIDS with an * denote clusters which were part of the B14 analysis.}\label{tab:data}}
\vspace{-2mm}
\begin{tabular}{lllllll}
\hline\hline
XCSID & $z_{\rm phot}^{\rm RM}$ & $T_{\textrm{X}}$ & $M_\text{DM}^\text{proj}$ & ObsID & $n_{\textrm{H}}$ \\
& & (keV) & (10$^{14}M_\odot$) & & (cm$^{-2}$) \\
\hline
XMMXCSJ000312.1-060530.5 & 0.251 & 6.81$_{-0.13}^{+0.22}$ & 6.52 & 0652010401 & 0.012 \\
XMMXCSJ000349.3+020404.8 & 0.11 & 4.78$_{-0.12}^{+0.12}$ & 3.8 & 0201900101 & 0.01 & \\
XMMXCSJ001053.4+290939.6 & 0.338 & 4.93$_{-0.37}^{+0.38}$ & 3.59 & 0650380101 & 0.024 \\
XMMXCSJ001737.5-005234.2 & 0.219 & 4.1$_{-0.22}^{+0.22}$ & 2.78 & 0403760701 & 0.022 \\
XMMXCSJ001833.2+162609.9 & 0.562 & 9.66$_{-0.36}^{+0.37}$ & 10.03 & 0111000101 & 0.021 \\
XMMXCSJ001938.0+033635.3 & 0.273 & 6.26$_{-0.15}^{+0.15}$ & 5.59 & 0693010301 & 0.035 \\
XMMXCSJ002635.9+170930.7 & 0.394 & 3.43$_{-0.14}^{+0.19}$ & 1.87 & 0050140201 & 0.03 \\
XMMXCSJ003456.6+023357.9 & 0.379 & 5.53$_{-0.41}^{+0.53}$ & 4.27 & 0650380601 & 0.017 \\
XMMXCSJ003706.4+090925.8 & 0.264 & 8.24$_{-0.26}^{+0.26}$ & 8.98 & 0084230201 & 0.049 \\
XMMXCSJ004630.7+202803.6 & 0.105 & 2.44$_{-0.23}^{+0.23}$ & 1.21 & 0652460101 & 0.031 \\
XMMXCSJ005138.5+271958.8 & 0.38 & 6.83$_{-0.32}^{+0.42}$ & 6.13 & 0650380701 & 0.02 \\
XMMXCSJ005559.1+261949.0 & 0.196 & 5.84$_{-0.16}^{+0.16}$ & 5.15 & 0203220101 & 0.028 \\
XMMXCSJ010649.3+010324.7 & 0.25 & 2.89$_{-0.04}^{+0.04}$ & 1.51 & 0762870601 & 0.036 \\
XMMXCSJ013724.6-082727.6 & 0.557 & 7.87$_{-0.66}^{+0.67}$ & 7.08 & 0700180201 & 0.01 \\
XMMXCSJ014656.7-092940.5 & 0.429 & 5.09$_{-0.35}^{+0.36}$ & 3.6 & 0673750101 & 0.029 \\
XMMXCSJ015242.1+010029.4 & 0.231 & 5.38$_{-0.15}^{+0.31}$ & 4.4 & 0084230401 & 0.029 \\
XMMXCSJ015334.1-011816.1 & 0.245 & 5.05$_{-0.19}^{+0.19}$ & 3.93 & 0762870401 & 0.029 \\
XMMXCSJ015707.7-055233.7 & 0.132 & 4.09$_{-0.23}^{+0.24}$ & 2.89 & 0781200101 & 0.032 \\
XMMXCSJ015824.9-014654.3 & 0.157 & 2.74$_{-0.11}^{+0.11}$ & 1.44 & 0762870301 & 0.008 \\
XMMXCSJ020143.0-021146.5 & 0.198 & 3.55$_{-0.08}^{+0.08}$ & 2.19 & 0605000301 & 0.022 \\
XMMXCSJ021441.2-043313.8 & 0.143 & 5.25$_{-0.22}^{+0.25}$ & 4.4 & 0553911401 & 0.02 \\
XMMXCSJ022145.6-034613.7 & 0.422 & 4.84$_{-0.41}^{+0.41}$ & 3.33 & 0604280101 & 0.009 \\
XMMXCSJ023142.5-045254.5 & 0.194 & 4.41$_{-0.18}^{+0.17}$ & 3.19 & 0762870201 & 0.023 \\
XMMXCSJ023953.0-013441.1 & 0.358 & 5.91$_{-0.16}^{+0.16}$ & 4.84 & 0782150101 & 0.043 \\
XMMXCSJ024803.3-033143.4* & 0.195 & 3.78$_{-0.06}^{+0.06}$ & 2.45 & 0084230501 & 0.011 \\
XMMXCSJ024811.9-021624.9 & 0.241 & 7.74$_{-0.36}^{+0.36}$ & 8.15 & 0721890401 & 0.033 \\
XMMXCSJ025632.9+000558.5 & 0.364 & 4.9$_{-0.12}^{+0.12}$ & 3.51 & 0801610101 & 0.016 \\
XMMXCSJ073220.2+313751.1 & 0.182 & 5.94$_{-0.16}^{+0.16}$ & 5.34 & 0673850201 & 0.017 \\
XMMXCSJ080056.7+360323.0 & 0.292 & 5.93$_{-0.21}^{+0.21}$ & 5.03 & 0781590201 & 0.017 \\
XMMXCSJ082318.4+155758.0 & 0.159 & 3.01$_{-0.19}^{+0.2}$ & 1.69 & 0742510401 & 0.029 \\
XMMXCSJ082557.4+041445.6 & 0.238 & 4.65$_{-0.27}^{+0.28}$ & 3.42 & 0762950301 & 0.049 \\
XMMXCSJ085026.7+001506.2 & 0.201 & 3.21$_{-0.18}^{+0.18}$ & 1.84 & 0761730501 & 0.051 \\
XMMXCSJ085612.8+375605.7 & 0.401 & 5.42$_{-0.31}^{+0.52}$ & 4.08 & 0302581801 & 0.014 \\
XMMXCSJ090036.8+205340.6 & 0.244 & 3.91$_{-0.09}^{+0.09}$ & 2.54 & 0402250701 & 0.03 \\
XMMXCSJ090849.1+143831.6 & 0.442 & 3.34$_{-0.17}^{+0.21}$ & 1.74 & 0674370201 & 0.048 \\
XMMXCSJ090851.4+144550.0 & 0.457 & 5.32$_{-0.42}^{+0.45}$ & 3.84 & 0674370201 & 0.037 \\
XMMXCSJ090912.4+105831.2 & 0.176 & 5.38$_{-0.16}^{+0.26}$ & 4.52 & 0673850901 & 0.017 \\
XMMXCSJ091048.8+385007.5 & 0.564 & 9.55$_{-0.78}^{+0.78}$ & 9.81 & 0723780101 & 0.032 \\
XMMXCSJ091110.7+174627.4 & 0.514 & 6.61$_{-0.3}^{+0.33}$ & 5.38 & 0693662501 & 0.019 \\
XMMXCSJ091345.5+405626.3 & 0.424 & 5.94$_{-0.25}^{+0.25}$ & 4.71 & 0147671001 & 0.02 \\
XMMXCSJ091752.2+514332.6* & 0.228 & 7.25$_{-0.2}^{+0.2}$ & 7.35 & 0084230601 & 0.008 \\
XMMXCSJ092018.6+370622.2 & 0.239 & 2.63$_{-0.05}^{+0.05}$ & 1.29 & 0149010201 & 0.014 \\
XMMXCSJ094300.0+465937.3 & 0.348 & 5.09$_{-0.18}^{+0.18}$ & 3.77 & 0106460101 & 0.045 \\
XMMXCSJ100304.6+325339.3 & 0.391 & 3.17$_{-0.26}^{+0.26}$ & 1.64 & 0302581601 & 0.03 \\
XMMXCSJ100742.4+380046.1 & 0.106 & 3.24$_{-0.16}^{+0.16}$ & 1.96 & 0653450201 & 0.013 \\
XMMXCSJ101703.4+390250.1 & 0.208 & 6.11$_{-0.13}^{+0.13}$ & 5.54 & 0084230701 & 0.023 \\
XMMXCSJ102339.7+041115.3* & 0.291 & 5.4$_{-0.03}^{+0.03}$ & 4.3 & 0605540201 & 0.021 \\
XMMXCSJ103801.2+414619.8 & 0.133 & 2.07$_{-0.15}^{+0.21}$ & 0.9 & 0206180101 & 0.016 \\
XMMXCSJ104044.2+395711.1* & 0.142 & 3.79$_{-0.05}^{+0.05}$ & 2.52 & 0147630101 & 0.019 \\
XMMXCSJ104545.6+042025.4 & 0.15 & 2.87$_{-0.21}^{+0.25}$ & 1.57 & 0653450601 & 0.034 \\
XMMXCSJ104724.0+151436.0 & 0.214 & 3.82$_{-0.3}^{+0.3}$ & 2.47 & 0721880101 & 0.007 \\
XMMXCSJ111253.4+132640.2* & 0.181 & 4.78$_{-0.08}^{+0.08}$ & 3.68 & 0500760101 & 0.035 \\
XMMXCSJ113313.2+500838.5 & 0.367 & 4.73$_{-0.33}^{+0.33}$ & 3.29 & 0650382001 & 0.021 \\
XMMXCSJ114224.9+583134.7 & 0.326 & 7.75$_{-0.75}^{+0.75}$ & 7.82 & 0650382201 & 0.022 \\
XMMXCSJ114935.6+222401.8 & 0.529 & 8.55$_{-0.55}^{+0.76}$ & 8.29 & 0693661701 & 0.018 \\
XMMXCSJ115518.2+232424.3 & 0.135 & 6.31$_{-0.07}^{+0.07}$ & 6.06 & 0551280201 & 0.024 \\
XMMXCSJ115827.8+262943.4 & 0.141 & 1.68$_{-0.05}^{+0.2}$ & 0.63 & 0601260201 & 0.014 \\
XMMXCSJ120022.7+032007.4 & 0.138 & 5.94$_{-0.12}^{+0.12}$ & 5.45 & 0827010301 & 0.006 \\
XMMXCSJ121937.0-031840.9 & 0.295 & 4.75$_{-0.35}^{+0.35}$ & 3.45 & 0693010401 & 0.015 \\
XMMXCSJ122656.3+334332.8 & 0.514 & 4.73$_{-0.32}^{+0.33}$ & 3.04 & 0200340101 & 0.022 \\

\hline
\end{tabular}
\end{center}
\end{table*}

\begin{table*}
\begin{center}
\contcaption{{}}
\vspace{-2mm}
\begin{tabular}{lllllll}
\hline\hline
XCSID & $z_{\rm phot}^{\rm RM}$ & $T_{\textrm{X}}$ & $M_\text{DM}^\text{proj}$ & ObsID & $n_{\textrm{H}}$ \\
& & (keV) & (10$^{14}M_\odot$) & & (cm$^{-2}$) \\
\hline
XMMXCSJ123355.5+152608.2 & 0.23 & 5.19$_{-0.23}^{+0.23}$ & 4.15 & 0404120101 & 0.029 \\
XMMXCSJ123422.8+094718.7 & 0.239 & 4.26$_{-0.1}^{+0.16}$ & 2.94 & 0673851101 & 0.024 \\
XMMXCSJ123618.1+285901.9 & 0.222 & 3.33$_{-0.23}^{+0.35}$ & 1.94 & 0722660201 & 0.067 \\
XMMXCSJ123658.8+631117.9 & 0.3 & 6.43$_{-0.43}^{+0.45}$ & 5.77 & 0402250101 & 0.04 \\
XMMXCSJ124133.3+325023.7 & 0.352 & 5.56$_{-0.39}^{+0.47}$ & 4.37 & 0056020901 & 0.034 \\
XMMXCSJ124401.5+165347.3 & 0.542 & 4.2$_{-0.17}^{+0.22}$ & 2.44 & 0302581501 & 0.021 \\
XMMXCSJ130357.9+673055.2 & 0.222 & 3.8$_{-0.26}^{+0.26}$ & 2.43 & 0136000101 & 0.015 \\
XMMXCSJ130749.5+292549.3 & 0.261 & 3.11$_{-0.18}^{+0.18}$ & 1.7 & 0205910101 & 0.048 \\
XMMXCSJ131129.8-012024.5* & 0.185 & 8.06$_{-0.08}^{+0.08}$ & 8.99 & 0093030101 & 0.018 \\
XMMXCSJ131145.1+220206.1 & 0.17 & 3.52$_{-0.27}^{+0.32}$ & 2.2 & 0402250301 & 0.046 \\
XMMXCSJ132250.7+313911.4 & 0.317 & 6.65$_{-0.32}^{+0.55}$ & 6.05 & 0650384601 & 0.012 \\
XMMXCSJ133048.5-015149.4 & 0.103 & 4.21$_{-0.07}^{+0.07}$ & 3.08 & 0112240301 & 0.018 \\
XMMXCSJ133108.4-014338.4 & 0.545 & 3.79$_{-0.25}^{+0.25}$ & 2.04 & 0112240301 & 0.024 \\
XMMXCSJ133233.8+502450.2 & 0.274 & 5.94$_{-0.35}^{+0.35}$ & 5.1 & 0142860201 & 0.039 \\
XMMXCSJ133244.2+503243.5 & 0.286 & 7.24$_{-0.23}^{+0.22}$ & 7.11 & 0142860201 & 0.01 \\
XMMXCSJ133421.5+503058.9 & 0.585 & 4.62$_{-0.38}^{+0.39}$ & 2.8 & 0111160101 & 0.026 \\
XMMXCSJ133519.5+410004.9* & 0.234 & 7.16$_{-0.25}^{+0.25}$ & 7.16 & 0084230901 & 0.015 \\
XMMXCSJ133648.8+102624.0 & 0.159 & 3.1$_{-0.22}^{+0.22}$ & 1.78 & 0761590701 & 0.034 \\
XMMXCSJ140101.9+025238.3* & 0.253 & 6.52$_{-0.04}^{+0.04}$ & 6.04 & 0551830201 & 0.036 \\
XMMXCSJ141627.7+231523.5 & 0.137 & 3.28$_{-0.12}^{+0.12}$ & 1.98 & 0722140401 & 0.019 \\
XMMXCSJ141956.1+063434.9 & 0.541 & 4.23$_{-0.36}^{+0.43}$ & 2.47 & 0303670101 & 0.021 \\
XMMXCSJ142039.8+395505.8 & 0.575 & 8.1$_{-0.48}^{+0.48}$ & 7.36 & 0693661001 & 0.018 \\
XMMXCSJ142348.0+240444.1 & 0.523 & 5.63$_{-0.16}^{+0.16}$ & 4.08 & 0720700301 & 0.018 \\
XMMXCSJ142521.4+631143.1 & 0.14 & 4.86$_{-0.13}^{+0.13}$ & 3.87 & 0765031201 & 0.049 \\
XMMXCSJ142601.0+374937.0* & 0.175 & 8.3$_{-0.11}^{+0.11}$ & 9.5 & 0112230201 & 0.018 \\
XMMXCSJ143150.0+133159.5 & 0.166 & 3.63$_{-0.16}^{+0.16}$ & 2.32 & 0601970101 & 0.033 \\
XMMXCSJ144219.8+221809.9 & 0.107 & 3.49$_{-0.16}^{+0.11}$ & 2.23 & 0765010501 & 0.015 \\
XMMXCSJ145715.0+222032.3 & 0.267 & 4.47$_{-0.06}^{+0.06}$ & 3.15 & 0108670201 & 0.036 \\
XMMXCSJ150019.6+212214.5 & 0.162 & 5.78$_{-0.15}^{+0.15}$ & 5.15 & 0693011001 & 0.023 \\
XMMXCSJ150817.8+575437.8 & 0.55 & 8.36$_{-0.54}^{+0.72}$ & 7.88 & 0723780501 & 0.023 \\
XMMXCSJ151012.0+333058.0* & 0.121 & 6.34$_{-0.15}^{+0.15}$ & 6.14 & 0149880101 & 0.025 \\
XMMXCSJ151618.5+000532.4 & 0.12 & 4.69$_{-0.09}^{+0.09}$ & 3.66 & 0201902001 & 0.021 \\
XMMXCSJ151820.6+292735.3 & 0.558 & 6.45$_{-0.25}^{+0.25}$ & 5.04 & 0693661101 & 0.023 \\
XMMXCSJ152642.6+164734.9 & 0.341 & 4.47$_{-0.27}^{+0.28}$ & 3.03 & 0650382801 & 0.018 \\
XMMXCSJ152925.0+104144.0 & 0.488 & 5.01$_{-0.16}^{+0.16}$ & 3.4 & 0762520201 & 0.019 \\
XMMXCSJ153253.8+302100.5* & 0.357 & 5.03$_{-0.08}^{+0.08}$ & 3.67 & 0651240101 & 0.008 \\
XMMXCSJ153941.0+342512.8 & 0.236 & 6.7$_{-0.27}^{+0.28}$ & 6.39 & 0673850601 & 0.046 \\
XMMXCSJ163936.8+470310.0 & 0.226 & 4.04$_{-0.33}^{+0.36}$ & 2.7 & 0761590401 & 0.032 \\
XMMXCSJ164020.2+464227.1 & 0.233 & 9.86$_{-0.3}^{+0.3}$ & 12.39 & 0605000501 & 0.023 \\
XMMXCSJ165943.9+323654.9 & 0.102 & 3.71$_{-0.31}^{+0.3}$ & 2.48 & 0083150801 & 0.012 \\
XMMXCSJ172227.0+320758.0 & 0.229 & 7.09$_{-0.14}^{+0.14}$ & 7.06 & 0693180901 & 0.019 \\
XMMXCSJ212939.7+000516.9* & 0.248 & 5.2$_{-0.06}^{+0.06}$ & 4.12 & 0093030201 & 0.033 \\
XMMXCSJ213516.8+012600.0 & 0.237 & 8.59$_{-0.32}^{+0.58}$ & 9.76 & 0692931301 & 0.017 \\
XMMXCSJ215101.0-073633.5 & 0.274 & 4.13$_{-0.12}^{+0.12}$ & 2.74 & 0744390301 & 0.021 \\
XMMXCSJ215337.0+174146.9* & 0.251 & 10.08$_{-0.25}^{+0.25}$ & 12.75 & 0111270101 & 0.02 \\
XMMXCSJ221145.8-034936.8 & 0.424 & 10.55$_{-0.24}^{+0.24}$ & 12.59 & 0693010601 & 0.016 \\
XMMXCSJ222353.0-013714.4 & 0.101 & 4.39$_{-0.1}^{+0.08}$ & 3.31 & 0401920101 & 0.016 \\
XMMXCSJ222605.0+172220.2 & 0.114 & 6.17$_{-0.08}^{+0.08}$ & 5.88 & 0762470101 & 0.038 \\
XMMXCSJ222831.6+203729.9 & 0.413 & 8.09$_{-0.26}^{+0.26}$ & 8.04 & 0147890101 & 0.044 \\
XMMXCSJ224321.4-093550.2 & 0.435 & 6.77$_{-0.09}^{+0.17}$ & 5.86 & 0503490201 & 0.015 \\
XMMXCSJ224413.0-093427.9 & 0.444 & 3.45$_{-0.24}^{+0.33}$ & 1.84 & 0503490201 & 0.04 \\
XMMXCSJ224523.7+280802.8 & 0.346 & 5.63$_{-0.51}^{+0.52}$ & 4.48 & 0650384401 & 0.05 \\
XMMXCSJ230821.8-021127.4 & 0.3 & 7.81$_{-0.49}^{+0.49}$ & 8.04 & 0205330501 & 0.021 \\
XMMXCSJ231132.6+033759.9 & 0.304 & 6.55$_{-0.27}^{+0.27}$ & 5.93 & 0693010101 & 0.044 \\
XMMXCSJ231825.4+184246.9 & 0.163 & 3.33$_{-0.11}^{+0.11}$ & 2.0 & 0762950201 & 0.039 \\
XMMXCSJ233738.6+001614.5 & 0.295 & 7.21$_{-0.36}^{+0.36}$ & 7.02 & 0042341301 & 0.02 \\
XMMXCSJ234116.6-090128.8 & 0.258 & 6.77$_{-0.24}^{+0.35}$ & 6.43 & 0693010801 & 0.023 \\
\hline
\end{tabular}
\end{center}
\end{table*}

\section{MOS data for individual clusters}

\begin{figure}
    \centering
    \includegraphics[width=0.45\textwidth]{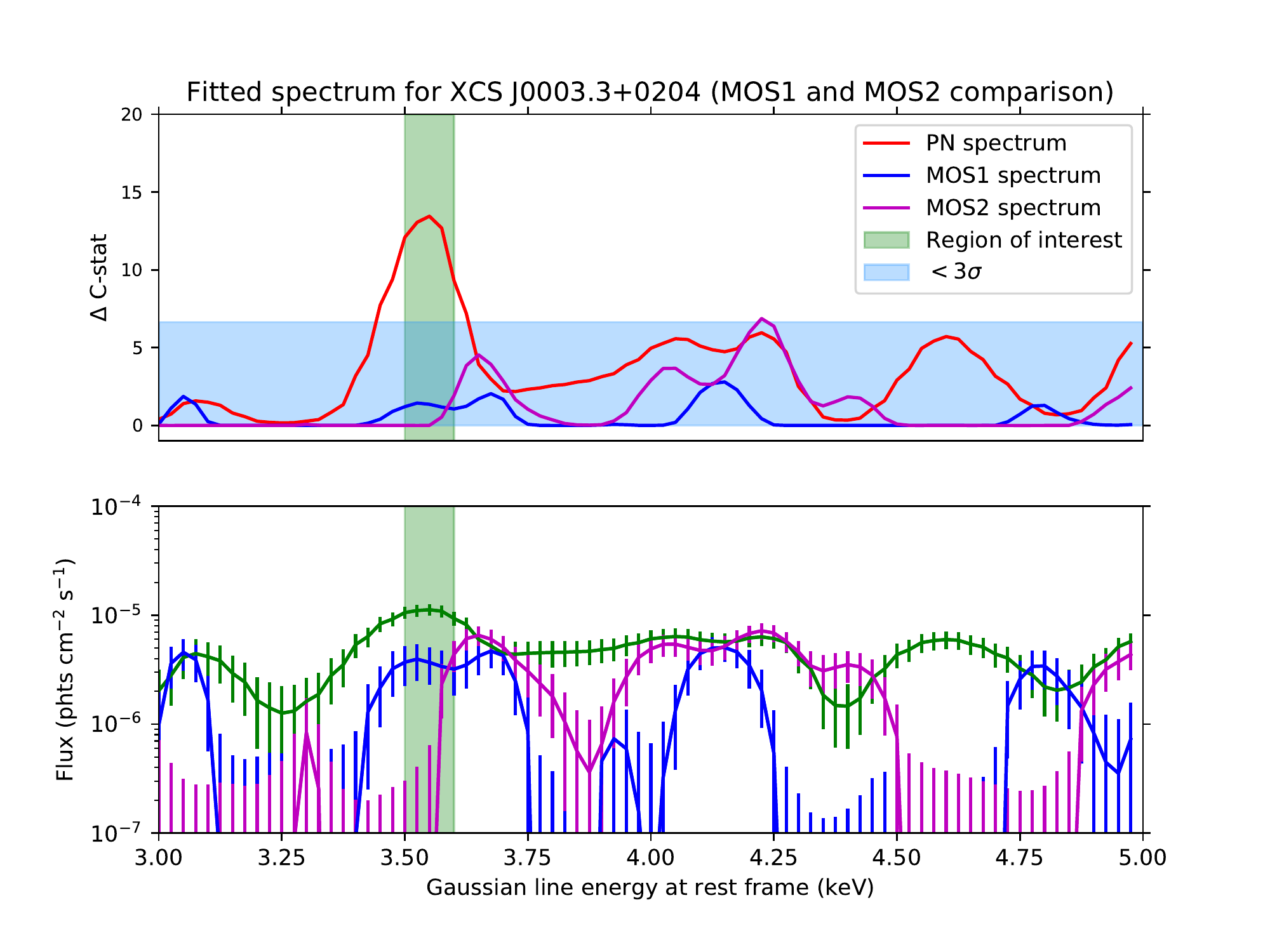} \\
    (a)\\
    \includegraphics[width=0.45\textwidth]{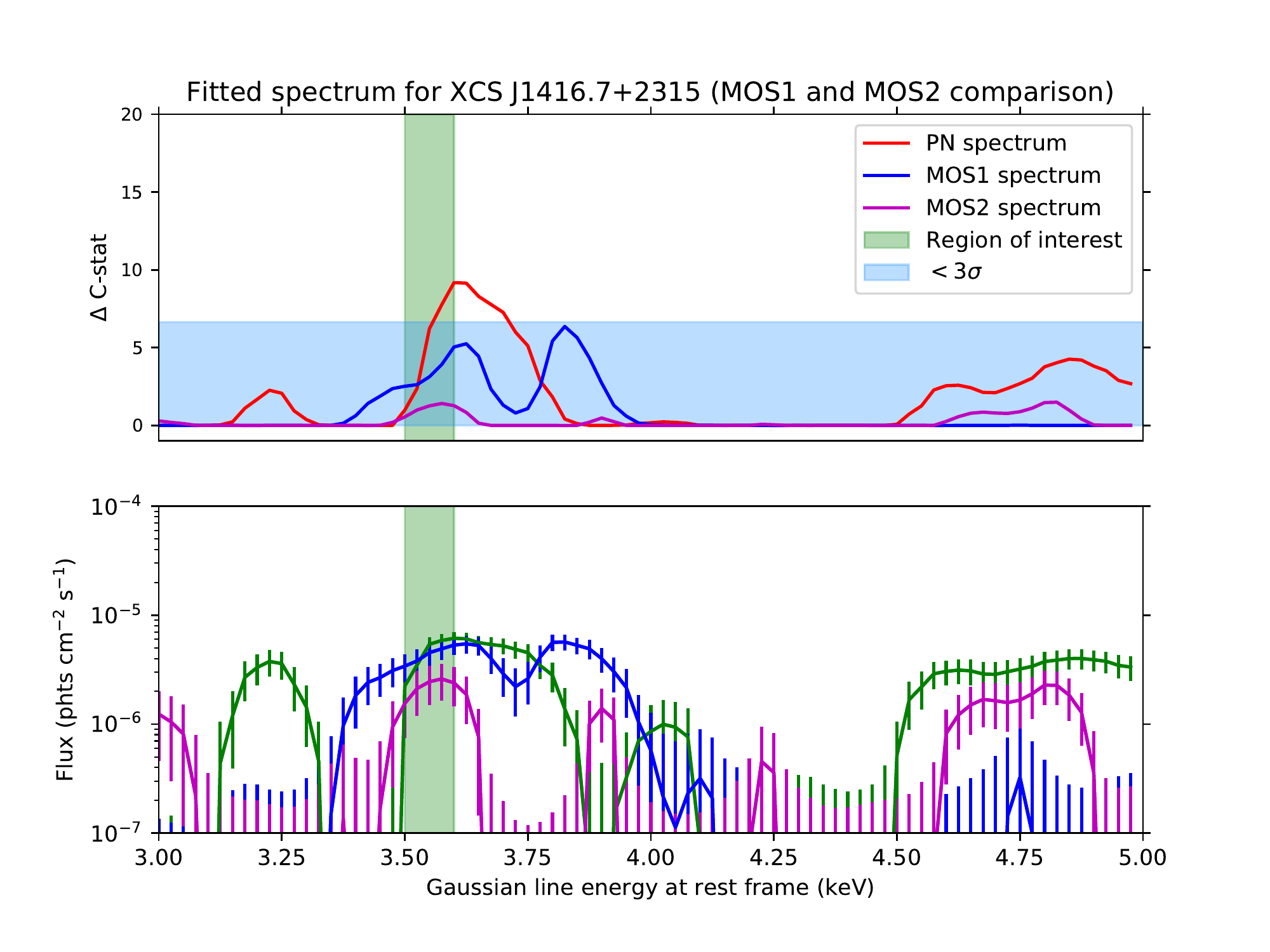} \\
    (b)\\
    \includegraphics[width=0.45\textwidth]{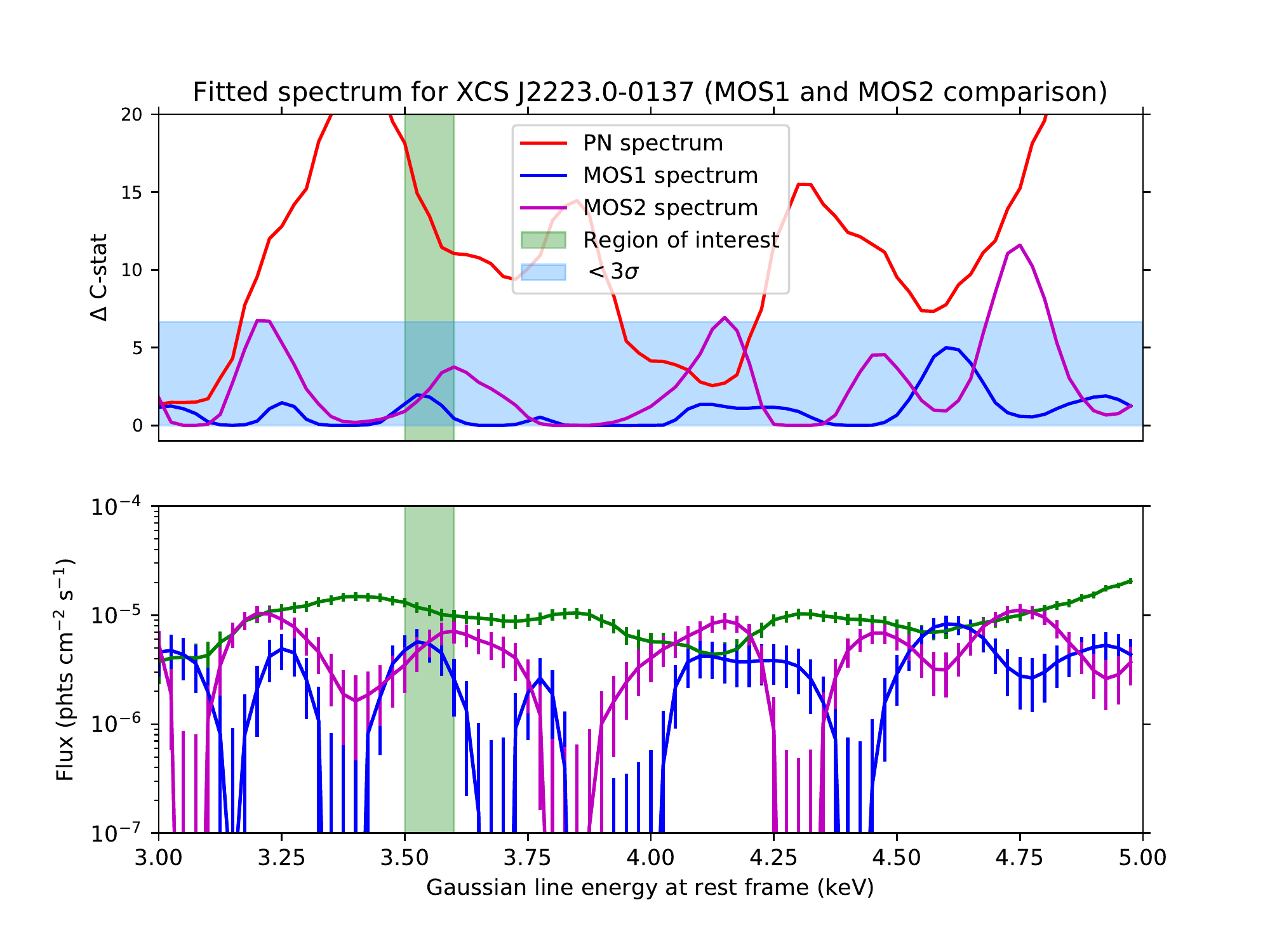} \\
    (c)\\
    \caption{Comparisons in the trend of $\Delta C$ for (a) XCS J0003.3+0204, (b) XCSJ1416.7+2315 and (c) XCS J2223.0-0137 using the highest quality PN and MOS observation for each cluster (described in Section~\ref{sec:clustercontributions}). In the top panels, the $\Delta C$ trend is displayed for the PN (red), MOS1 (blue) and MOS2 (magenta) spectra. In the bottom panels, the corresponding Gaussian normalisation and associated errorbars are shown.}
    \label{fig:moscomparisons}
\end{figure}

Figure \ref{fig:moscomparisons} shows the corresponding trends in $\Delta C$ for the three individual clusters in the analysis using available MOS data.

\section{Spectroscopic redshift comparisons}

\begin{figure}
    \centering
    \includegraphics[width=0.45\textwidth]{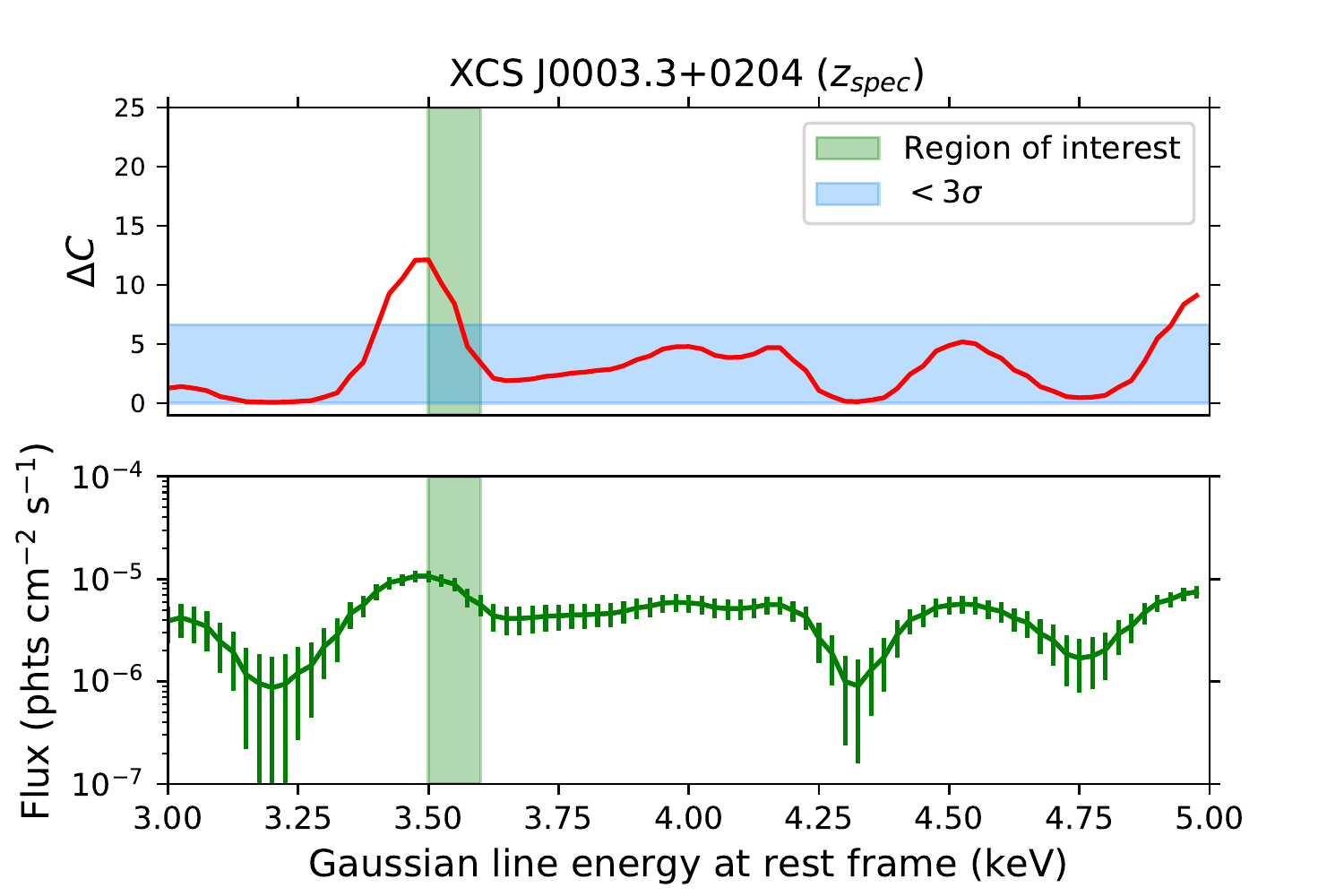}
    \includegraphics[width=0.45\textwidth]{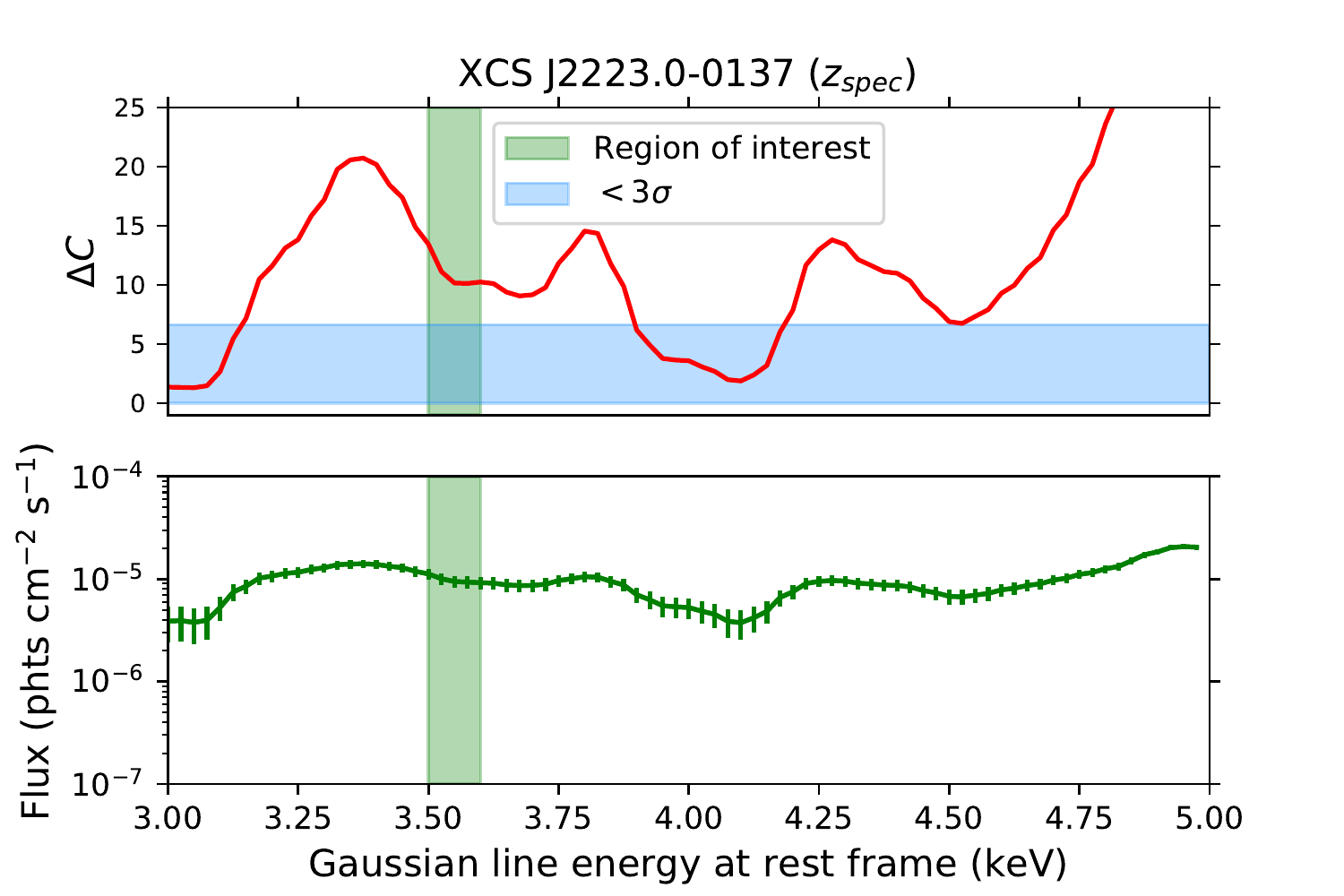}
    \caption{Comparisons in the trend of $\Delta C$ for XCS J0003.3+0204 (top) and XCS J2223.0-0137 (bottom) when replacing the RM photometric estimated redshift with available spectroscopic redshifts (see Sect.~\ref{subsec:photo-z}).}
    \label{fig:spectrozs}
\end{figure}

Figure \ref{fig:spectrozs} shows the comparison in the $\Delta C$ trend for two clusters using available spectroscopic data.

\bsp	
\label{lastpage}
\end{document}